\documentclass[tighten,trackchanges]{aastex62}

\newcommand{\bOmega}{\bar{\Omega}}

\usepackage{xspace}

\newcommand*\diff{\mathop{}\!\mathrm{d}}

\renewcommand*{\deg}{\degr{} } 
\newcommand*{\msol}{M$_\odot$\xspace} 

\submitjournal{Astrophysical Journal Letters}

\shorttitle{Emission from a Rapidly Rotating Neutron Star}
\shortauthors{Bogdanov et al.}

\begin{document}

\title{CONSTRAINING THE NEUTRON STAR MASS-RADIUS RELATION AND DENSE MATTER EQUATION OF STATE WITH \textsl{NICER}. II. EMISSION FROM HOT SPOTS ON A RAPIDLY ROTATING NEUTRON STAR}

\correspondingauthor{Slavko Bogdanov}
\email{slavko@astro.columbia.edu}

\author[0000-0002-9870-2742]{Slavko Bogdanov}
\affil{Columbia Astrophysics Laboratory, Columbia University, 550 West 120th Street, New York, NY 10027, USA}

\author[0000-0002-3862-7402]{Frederick K.~Lamb}
\affil{Center for Theoretical Astrophysics and Department of Physics, University of Illinois at Urbana-Champaign, 1110 West Green Street, Urbana, IL 61801-3080, USA}
\affil{Department of Astronomy, University of Illinois at Urbana-Champaign, 1002 West Green Street, Urbana, IL 61801-3074, USA}

\author{Simin Mahmoodifar}
\affiliation{Astrophysics Science Division and Joint Space-Science Institute, NASA Goddard Space Flight Center, Greenbelt, MD 20771, USA}

\author[0000-0002-2666-728X]{M.~Coleman Miller}
\affiliation{Department of 
Astronomy and Joint Space-Science Institute, University of Maryland, 
College Park, MD 20742-2421, USA}

\author[0000-0003-4357-0575]{Sharon M.~Morsink}
\affiliation{Department of Physics, University of Alberta, Edmonton, AB T6G 2G7, Canada}

\author[0000-0001-9313-0493]{Thomas E.~Riley}
\affiliation{Anton Pannekoek Institute for Astronomy, University of Amsterdam, Science Park 904, 1090GE Amsterdam, the Netherlands}

\author[0000-0001-7681-5845]{Tod E.~Strohmayer}
\affiliation{Astrophysics Science Division and Joint Space-Science Institute, NASA Goddard Space Flight Center, Greenbelt, MD 20771, USA}

\author{Albert K.~Tung}
\affiliation{Department of Physics, University of Alberta, Edmonton, AB T6G 2G7, Canada}

\author[0000-0002-1009-2354]{Anna L.~Watts}
\affiliation{Anton Pannekoek Institute for Astronomy, University of Amsterdam, Science Park 904, 1090GE Amsterdam, the Netherlands}

\author{Alexander J. Dittmann}
\affil{Department of 
Astronomy and Joint Space-Science Institute, University of Maryland, 
College Park, MD 20742-2421, USA}

\author[0000-0001-8804-8946]{Deepto Chakrabarty}
\affil{MIT Kavli Institute for Astrophysics and Space Research, Massachusetts Institute of Technology, 70 Vassar Street, Cambridge, MA 02139, USA}

\author[0000-0002-6449-106X]{Sebastien Guillot}
\affil{IRAP, CNRS, 9 avenue du Colonel Roche, BP 44346, F-31028 Toulouse Cedex 4, France}
\affil{Universit\'{e} de Toulouse, CNES, UPS-OMP, F-31028 Toulouse, France}

\author{Zaven Arzoumanian}
\affiliation{X-Ray Astrophysics Laboratory, NASA Goddard Space Flight Center, Greenbelt, MD 20771, USA}

\author{Keith C.~Gendreau} 
\affiliation{X-Ray Astrophysics Laboratory, NASA Goddard Space Flight Center, Greenbelt, MD 20771, USA}

\begin{abstract}
We describe the model of surface emission from a rapidly rotating neutron star that is applied to \textsl{Neutron Star Interior Composition Explorer} X-ray data of millisecond pulsars in order to statistically constrain the neutron star mass-radius relation and dense matter equation of state. 
To ensure that the associated calculations are both accurate and precise, we conduct an extensive suite of verification tests between our numerical codes for both the Schwarzschild + Doppler and Oblate Schwarzschild approximations, and compare both approximations against exact numerical calculations. We find superb agreement between the code outputs, as well as in comparison against a set of analytical and semi-analytical calculations, which combined with their speed, demonstrates that the codes are well-suited for large-scale statistical sampling applications. A set of verified, high-precision reference synthetic pulse profiles is provided to the community to facilitate testing of other independently developed codes.
\end{abstract}


\section{Introduction} \label{sec:intro}
If a spinning neutron star (NS) radiates X-rays from one or more hotter regions (hereafter ``hot spots'') on or near its surface and the gas in these spots rotates with the star at a regular rate, a distant observer will see periodic, energy-dependent pulsations. It is widely recognized that the observed pulsations offer a valuable  probe of the physical conditions and processes occurring at or very near the neutron star surface. Statistical estimates of the gravitational mass $M$ and the circumferential radius $R$ of several NSs with sufficiently different masses are of particular interest: they can provide crucial information about the formation and evolution of NSs, and can be used to constrain the properties of the cold, dense interior matter (see, e.g., \citealt{2016ApJ...820...28O,2016EPJA...52...18S,2019arXiv190501081B}). The stellar properties $M$ and $R$ can be jointly estimated by fitting models to the observed X-ray pulsations, because the properties of the observed pulsed signal are affected by both parameters. The characteristics (or absence) of occultations of the hotter regions as the star rotates constrain the stellar radius. The observed morphology of the pulsations depends on $M$ via general relativistic light deflection, which increases with increasing stellar compactness $M/R$. For discussions of the various approximations that have been used to compute these effects, see \citet{1983ApJ...274..846P,1992ApJ...388..138S,1998ApJ...499L..37M,2000ApJ...531..447B,2002ApJ...566L..85B,2003MNRAS.343.1301P,2007ApJ...654..458C,2007ApJ...663.1244M,2014ApJ...792...87P,2018A&A...615A..50N}.

The line-of-sight velocity of the X-ray emitting gas, and hence the relativistic Doppler boost and aberration it produces, is proportional to the product of the stellar radius and the spin rate. Hence, other things being equal, the changes in the pulse profile produced by these effects are larger for stars that are spinning more rapidly. Consequently, waveform fitting usually provides the strongest constraints on $R$ and $M$ when it is used to analyze the pulse profiles produced by NSs with millisecond spin periods. NSs that produce X-ray flux oscillations with millisecond periods include accretion-powered millisecond X-ray pulsars (see \citealt{2003MNRAS.343.1301P,2006MNRAS.373..836P,2008ApJ...672.1119L,2009ApJ...691.1235L,2011ApJ...742...17L,2011ApJ...726...56M,2018A&A...618A.161S} 
for recent models, and \citealt{2012arXiv1206.2727P} for a description of the observations); NSs that produce thermonuclear X-ray bursts that have millisecond brightness oscillations (see \citealt{1997ApJ...487L..77S,1998ApJ...499L..37M,2001ApJ...546.1098W,2005ApJ...619..483B,2013MNRAS.433L..64A,2013ApJ...776...19L,2014ApJ...787..136P,2015ApJ...808...31M,2015ApJ...811..144B} for pulse profile models, and \citealt{2012ARA&A..50..609W} for a discussion of the observations), and rotation-powered millisecond pulsars that appear to have hotter regions near their magnetic polar caps that produce periodic X-ray brightness modulations (see \citealt{2000ApJ...531..447B,2007ApJ...670..668B,2008ApJ...689..407B,2013ApJ...762...96B} for the models).

The observed pulse profile depends on $M$ and $R$ in various ways, both in the combination of the compactness ratio $M/R$ and separately. It follows that these parameters can in principle be measured individually by carefully analyzing the waveform of the X-ray emission from the star. Providing such measurements is one of the principal goals of NASA's \textsl{Neutron Star Interior Composition Explorer} (\textsl{NICER}) mission (see \citealt{2016SPIE.9905E..1HG}), which was launched and installed on the International Space Station in June 2017.  The first crucial step in assuring the correctness of such an analysis is to verify that the model pulse profiles being used are computed correctly for the assumed properties of the star, the assumed properties of its emission, and the direction and distance to the observer. This paper documents our procedures for performing such calculations.

The purpose of this paper is: i) to describe the model we apply to energy-resolved X-ray pulsations of millisecond pulsars observed by \textit{NICER} to obtain constraints on their $M$-$R$ relation; ii) present the results of a series of tests we have devised to ensure that our algorithms and codes produce precise and accurate results; iii) offer clarifications and corrections to the procedures commonly used to model the surface emission from rapidly rotating neutron stars; iv) provide a set of verified high-precision synthetic pulse profiles for those who wish to verify their own calculations.

This paper is the second in a series of papers dedicated to obtaining new information about the NS $M$ and $R$ and the dense matter equation of state using data of several nearby millisecond pulsars obtained with \textit{NICER}. In \citet[][Paper I hereafter]{bogdanov19a}, we present the data collected so far for target MSPs that are being analysed for this purpose.  \citet[][Paper III hereafter]{bogdanov19c} describes all other aspects of the modeling technique applied to the \textit{NICER} data, including neutron star atmospheres, interstellar absorption, the instrument response, and the $M$ and $R$ parameter estimation methodology, and the potential sources of systematic error. The first set of results for PSR J0030$+$0451 of the parameter estimation analyses that are based in part on the model described here are presented in \cite{miller19} and \cite{riley19}. 

The present paper is organized as follows.  In Section~\ref{sec:SD} we describe the essential ingredients of the Schwarzschild + Doppler (S+D) approximation, while in Section~\ref{sec:OS}  we explore the next-order Oblate Schwarzschild (OS) approximation \citep{2007ApJ...654..458C,2007ApJ...663.1244M}.  In Sections~\ref{sec:verification} and \ref{sec:ostests} we give the results from the comparison tests that have been performed on the S+D and OS approximations respectively between our codes as well as against exact general relativistic numerical calculations. In the appendices, we give examples of analytic or semi-analytic calculations that agree with the results of our codes and we present descriptions of the codes used in the verification. 

\begin{figure}[b]
\begin{center}
\includegraphics[clip, trim=3cm 3cm 3cm 3cm, width=0.7\textwidth]{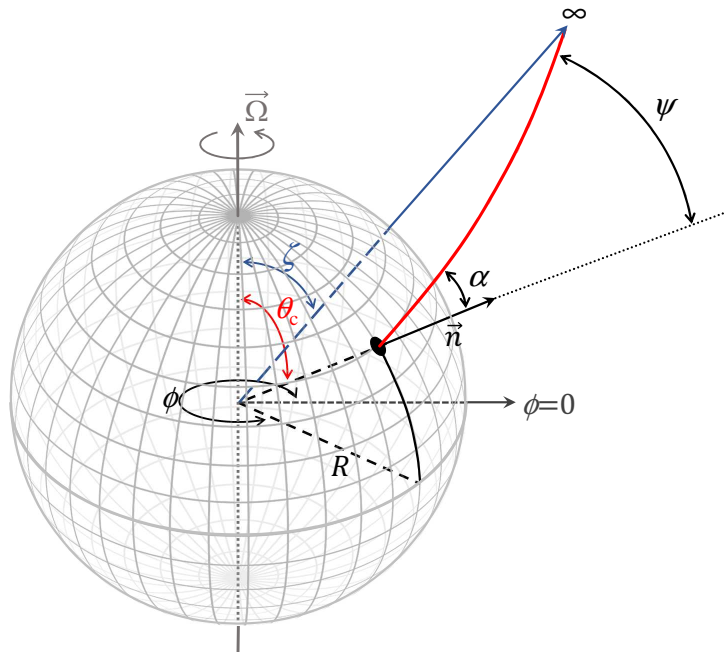}
\caption{\small{The geometry of a hot spot on the surface of a
    neutron star rotating at an angular frequency $\Omega$. A photon emitted from the
    surface at an angle $\alpha$ with respect to the local surface
    normal ${\vec n}$, as seen in the local static frame, deflects by a total angle $\psi$ as it travels to a
    distant observer. Figure adapted from \citet{2016EPJA...52...37B}.}}
\label{fig:nsdiagram}
\end{center}
\end{figure}

\section{Modeling Hot Spot Emission from Neutron Stars in the S+D Approximation}
\label{sec:SD}

A rotating NS is not perfectly spherical and the spacetime external to the star is not exactly Schwarzschild.  If we wished to compute the waveforms from rotating hot spots with the greatest possible accuracy and precision, we would therefore have to solve the exact general relativistic equations for the stellar shape and exterior spacetime metric using an axisymmetric code such as RNS \citep{1995ApJ...444..306S} or LORENE/NROTSTAR \citep{Vincent18} and then trace the rays in that spacetime.

Numerical relativistic raytracing using the metric corresponding to a rapidly rotating neutron star with a tabulated equation of state was first performed by \citet{2007ApJ...654..458C}. Their procedure involved first choosing an equation of state, and computing the equations of relativistic stellar structure using the RNS code for a choice of mass and spin. This results in a solution for the non-spherical shape of the rotating star and the metric for the star evaluated on a spatial grid. This calculation takes on the order of tens of seconds to compute on a modern processor. Once the metric has been computed, an initial latitude on the star's surface is chosen, and geodesics emitted from this location are emitted in all directions allowed by the star's surface. Once the geodesics connecting an initial and final angular location are found for all values of phase, a pulse waveform can be constructed. On a modern processor, this procedure takes on the order of many tens of minutes to construct a pulse profile.  Alternate codes 
\citep{2018A&A...615A..50N,Pihajoki2018,Vincent18}
that can do the same problem with similar accuracy and speed have been developed more recently.

The analysis of \textsl{NICER} data requires the generation of up to hundreds of millions of synthetic pulse waveforms.  Thus, speed as well as accuracy is essential.  With that in mind, the first approximation we will make is the S+D approximation (see \citealt{1998ApJ...499L..37M,2002ApJ...564..353N,2003MNRAS.343.1301P,2004AIPC..714..245S,2013ApJ...776...19L}).  In this approximation, all special-relativistic effects at the stellar surface are treated correctly but the star is approximated as a sphere and the spacetime external to the star is assumed to be the Schwarzschild spacetime.  
\citet{2007ApJ...654..458C} showed that their more general treatment of geodesics on the correct numerical spacetime reduces to the S+D approximation in the slow-rotation limit.
This approximation is extremely fast and is useful for slowly rotating stars (for rotation frequencies $\nu$ that are less than $\approx 100$~Hz), since we expect that observational data will have statistical errors larger than the waveform differences introduced by the rotational deformation of the metric and the surface. The more accurate OS approximation, appropriate for more rapidly spinning pulsars, will be investigated in more detail in the following section.

We begin in Subsection \ref{sec:rotating-spot} with a discussion of emission of light from a spot on a non-gravitating uniformly rotating sphere in the context of special relativity in order to point out an error in a number of previous publications when the surface area is defined in the comoving frame.  This is followed in Subsection~\ref{sec:gr} with the addition of gravity to the rotating sphere. Since in what follows we make use of a large number of variables, with symbols that are not consistently used in the literature, for convenience we provide Figure \ref{fig:nsdiagram} and Table \ref{table:parameters} in the Appendix summarizing the notation used here.

\subsection {Emission from the Surface of a Uniformly Rotating Non-Gravitating Sphere}
\label{sec:rotating-spot}

Consider the emission of light from a non-gravitating sphere which has a radius $R$ as measured in a local static frame and rotates uniformly with an angular frequency $\Omega$ as seen in that frame.  We wish to compute the energy-resolved flux seen by a distant observer from a spot that rotates with the sphere on its surface.  The observer is at a distance $D$ from the star, far enough away that the light rays originating from the sphere are parallel to each other.  There is no gravity, so light rays are straight lines.  The considerations we discuss will also apply to oblate stars, but the spherical case is easier to picture.

\subsubsection{Emission from an \textbf{Infinitesimal} Patch on the Surface of the Sphere}

Suppose that we are interested in an infinitesimal patch of the emitting star, which has a linear extent that corresponds to a range of colatitudes between $\theta$ and $\theta+{\rm d}\theta$, and to a range of azimuths between $\phi$ and $\phi+{\rm d}\phi$.  The star rotates in the $+\phi$ direction as seen by static observers.  The solid angle is a Lorentz invariant, which means that both a local comoving observer riding on the patch and a local static observer directly above the patch will measure the solid angle of the patch to be $\sin\theta\,{\rm d}\theta\,{\rm d}\phi$.  

However, as is shown in various references (e.g., the excellent pedagogical discussion in \citealt{2012AmJPh..80..772K}), although the linear extent of the patch in the $\theta$ direction, $R{\rm d}\theta$, is the same in the comoving and the static frames, the linear extent of the patch in the $\phi$ direction, which is measured to be $R\sin\theta\,{\rm d}\phi$ by a static observer directly above the patch, is measured to be $\gamma(\theta)R\sin\theta\,{\rm d}\phi$ by the comoving observer, where
\begin{equation}
\gamma(\theta)\equiv\left[1-\beta^2(\theta)\right]^{-1/2}
\label{eq:gamma}
\end{equation}
is the Lorentz factor corresponding to the dimensionless speed
\begin{equation}
\beta(\theta)\equiv \Omega R\sin\theta/c
\label{eq:beta}
\end{equation}
seen by a static observer at colatitude $\theta$.  Thus a spot that appears to be circular when measured in linear coordinates in the comoving frame will appear to be compressed in the direction of motion to a local static observer, while a star that appears spherical to local static observers will appear oblate to local comoving observers. 
This effect, derived purely in special relativity, also needs to be included (with modifications to the definition of the velocity) in a treatment with gravity.
Some previous publications \citep[e.g.,][]{1998ApJ...499L..37M,2003MNRAS.343.1301P,2007ApJ...670..668B,2008ApJ...672.1119L} used an incorrect expression for the comoving surface element, which neglected the factor $\gamma(\theta)$. This error was discussed by  \citet{psaltis_dimitrios_2019_3534254}, \citet{2018ApJ...854..187L}, and \citet{2018A&A...615A..50N}.
The codes used for the parameter estimation analyses presented in \cite{miller19} and \cite{riley19} correctly include this $\gamma(\theta)$ factor. 

Suppose now that we are interested in a light ray emerging from the star that makes an angle $\xi$ with the local direction of motion, as seen in the static frame.  Then, the Doppler factor
\begin{equation}
\delta\equiv {1\over{\gamma(\theta)[1-\beta(\theta)\cos\xi]}} \label{eq:delta}
\end{equation}
can be used to convert the values of several useful quantities from the comoving frame to the static frame.  Using the convention that primed quantities are measured in a local inertial frame that is momentarily comoving with the stellar surface (hereafter referred to as the local comoving frame) whereas unprimed quantities are measured in the local static frame, the angles from the local surface normal are related by
\begin{equation}
\cos\alpha=\delta^{-1}\cos\alpha^\prime\; ,
\end{equation}
photon energies are related by
\begin{equation}
E=\delta E^\prime\; ,
\label{eq:specE}
\end{equation}
and specific intensities are related by 
\begin{equation}
I(E,\alpha)=\delta^3 I^\prime(E^\prime,\alpha^\prime)\; .
\label{eq:conservation}
\end{equation}

An observer at a distance $D\gg R$ from the star will therefore measure the flux from an infinitesimal patch on the star that is centered at $(\theta^\prime,\phi^\prime)=(\theta,\phi)$ to be
\begin{equation}
\diff F(E,\alpha,\theta,\phi)=I(E,\alpha,\theta,\phi){\diff S\cos\alpha\over{D^2}}=\delta^3I^\prime(E^\prime,\alpha^\prime,\theta^\prime,\phi^\prime){\diff S^\prime\cos\alpha^\prime\over{D^2}}\; ,
\end{equation}
where the surface area element in the local comoving frame is defined by
\begin{equation}
\diff S^\prime=\gamma(\theta^\prime)R^2\sin\theta^\prime \diff\theta^\prime \diff\phi^\prime 
= \gamma(\theta)R^2\sin\theta \diff \theta \diff \phi.
\end{equation}

The light travel time from the patch to the observer depends on the location of the patch at the time of emission.  One way to take this into account is to first compute the {\it non-rotating} azimuthal radiation pattern that would be seen at the inclination and distance of the observer, taking into account exactly all the special relativistic effects on the energy, beaming, etc., of the radiation that are produced by the star's rotation. Once this time-independent radiation pattern has been computed as a function of photon energy, the pattern can be rotated at the stellar spin frequency to obtain the observed time-dependent waveform.


Using this approach, the time-dependent flux at an observer located at $(\zeta, \phi_{\rm obs})$ and $\phi(t) = \Omega t$ can be written
\begin{equation}
{\rm d}F_{\rm obs}(\zeta, \phi_{\rm obs}, t) = {\rm d}F_0(\zeta, \phi(t))\; . 
\end{equation}
%
%
%
In this approach, it is the time-dependence of the argument $\phi(t)$ that produces the time dependence of the observed flux ${\rm d}F_{\rm obs}(\zeta, \phi_{\rm obs}, t)$.  Note that, as measured in the local comoving frame, the emission from the surface is steady (even variation produced by, e.g., spreading of thermonuclear burning occurs on time scales much longer than typical stellar rotation periods).  It is only the non-axisymmetry of the emission combined with rotation that causes the flux seen by a distant observer to vary.

\subsubsection{Emission from an Extended Spot on a Rotating Sphere}
\label{sec:rotating-sphere}

The flux as a function of time from an extended spot is simply the integral of the fluxes from the infinitesimal patches that make up the spot. Care must be taken to ensure that the photons that are counted in the flux all arrive at the distant observer at the same time.  That is, if the times needed to reach the distant observer from two different points of emission on the star differ by $\Delta t$, then the difference in the emission times from the two points needs to be $-\Delta t$ so that both of the rays reach the observer simultaneously.

As a specific example, consider a small spot on the equator of the rotating star, as seen by an equatorial observer.  Let $\phi=0$ be the point directly underneath the observer.  Then, emission from an infinitesimal patch at azimuth $\phi$ (where $-\pi/2\leq\phi\leq\pi/2$ because we assume a very distant observer and light travels on straight-line paths in our special-relativistic example) will have a propagation time to the observer that is $\Delta t=(R/c)(1-\cos\phi)$ longer than the propagation time of a radial ray.  Thus, if the pulse waveform is folded on the angular frequency $\Omega$, the flux from this emission will contribute to a phase $\phi_{\rm obs}=\phi+\Omega\Delta t$ rather than to $\phi_{\rm obs}=\phi$.

\subsection{Modeling Emission from a Gravitating Star in a Schwarzschild Exterior Spacetime}
\label{sec:gr}

We now move from consideration of rotating non-gravitating stars to rotating gravitating stars. Nothing about the transformation between local comoving and local static observers will change.  However, light will not travel in straight lines, and gravitational redshifts must be taken into account. 

A major advantage to using the Schwarzschild spacetime, in contrast to spacetimes that have frame-dragging, is that the spherical symmetry of the Schwarzschild geometry guarantees that the path followed by any given photon lies in a plane. Hence, the procedure for tracing the path of a light ray from any angular location $(\theta,\phi)$ on the stellar surface to any angular location $(\zeta,\phi_{\rm obs})$ at a large distance is simple:

\begin{enumerate}

\item Determine the deflection angle $\psi=\cos^{-1}\left[(\theta,\phi)\cdot(\zeta,\phi_{\rm obs})\right]$ between the starting and ending points of the ray.

\item Determine the angle $\alpha$ from the surface normal, as seen in the local static frame at the stellar surface, such that a photon leaving the surface at that angle will be deflected by an angle $\psi$ in propagating to infinity.J. Nättilä 

\item Using, e.g., spherical triangles (or the great circle distance), determine the local azimuthal angle $\lambda$ (defined, for example, so that $\lambda=0$ points north and $\lambda=\pi/2$ points east) such that an arc of angular size $\psi$, in the direction $\lambda$, connects $(\theta,\phi)$ at the surface to $(\zeta,\phi_{\rm dist})$ at infinity.

\end{enumerate}

Thus, no actual tracing of a photon is required: a simple table lookup of $\psi(\alpha)$ suffices to determine the needed direction from the stellar surface.  This results in a tremendous saving of computational time.

We show the geometry of the system in Figure~\ref{fig:nsdiagram}.  We consider an infinitesimal hot spot at a colatitude $\theta_c$ with respect to the stellar rotational pole (finite-sized hot spots can be built by linear addition of infinitesimal spots) seen by an observer at a colatitude $\zeta$.  If we denote by $\phi(t)$ the azimuthal angle of the spot as a function of time (where $\phi=0$ means that the spot is at the same longitude as the observer), then the angular distance $\psi(t)$ between the spot and the observer is given by
\begin{equation}
\cos\psi(t)=\sin \theta_c \sin \zeta \cos \phi (t) + \cos \theta_c \cos \zeta\; .
\end{equation}
It is useful to break the computation of the waveform seen by a distant observer into two separate frame shifts: (1)~from a surface comoving frame to a local static frame at the surface; and (2)~from the local static frame to the distant observer.  The first involves only local special relativistic transformations, whereas the second requires non-local general relativistic effects. In both cases, it is helpful to use the constancy of $I_E/E^3$ along rays, where $E$ is the energy of a photon and $I_E$ is the specific intensity at $E$, to follow the effect of redshifts and blueshifts on the specific intensity.  We now discuss the special relativistic effects, followed by the general relativistic effects.

\subsubsection{From the Surface Comoving Frame to the Static Frame}
\label{sec:sr}

It is typically assumed that an observer at the stellar surface who moves with the star will see a particularly simple specific intensity of emission.  For example, it is standard to assume that in this frame, the specific intensity depends only on the angle from the local normal and not on the azimuthal angle of emission. For testing purposes, we sometimes assume an isotropic beaming pattern, although such a pattern is not expected for real systems. For example, the non-accreting  rotation-powered NSs that are the focus of the \textsl{NICER} mission likely have the beaming patterns of non-magnetic light-element atmospheres (H or He; see \citealt{potekhin14} and references therein).   We use primes to denote quantities measured in the surface comoving frame, e.g., $\alpha^\prime$ is the angle from the normal as seen in the comoving frame.

The Lorentz transformation from the surface comoving frame to the frame of a temporarily co-located static observer is local. The speed of the surface as measured by that local static observer is
\begin{equation}
v=\Omega R(1-R_S/R)^{-1/2}\sin\theta_c\; , \label{eq:v}
\end{equation}
where $\Omega\equiv 2\pi\nu$ is the rotational angular frequency of the star as seen at infinity and $R_S=2GM/c^2$ is the Schwarzschild radius. The equations defining the transformation from the surface comoving frame to the local static frame are similar to those given in Section \ref{sec:rotating-sphere}, except that the velocity is given by equation (\ref{eq:v}), which is corrected for gravitational redshift. The relativistic Doppler factor $\delta$ is
\begin{equation}
\delta={1\over{\gamma[1-(v/c)\cos\xi}]}\; ,
\end{equation}
where $\gamma=[1-(v/c)^2]^{-1/2}$ is the Lorentz factor and $\xi$ is the angle of the photon's propagation relative to the direction of rotation, as seen by a local static observer. The relation between the angle $\alpha^\prime$ of the photon propagation direction relative to the surface normal in the comoving frame and the angle $\alpha$ relative to the surface normal in the local static frame is
\begin{equation}
\cos\alpha^\prime=\delta\cos\alpha\; .
\label{eq:alphatrans}
\end{equation}
Moreover, if the area of the infinitesimal surface patch is $dS^\prime$ as seen in the comoving frame, then the area of the same patch as seen in the local static frame is
\begin{equation}
\diff S=\delta \diff S^\prime \;,
\end{equation}
which means that $\diff S\cos\alpha=\diff S^\prime\cos\alpha^\prime$ is a Lorentz invariant \citep{1975pbrg.book.....L,1985ApJ...295..358L}.

\subsubsection{From the Surface Static Frame to the Distant Observer}
\label{sec:gr1}
Once the transformation has been made to the surface static frame, the propagation of photons to a distant observer introduces several effects:

{\it Gravitational redshift.}---In the Schwarzschild spacetime, gravitational redshifts depend only on the initial and final radius and not on the direction of propagation.  The photon energy seen by an observer at a large distance $D$ is $(1-R_S/R)^{1/2}$ times the photon energy seen by an observer in the local static frame at the stellar radius $R$. The relation between the photon energy $E'$  emitted in the local comoving frame and the photon energy $E$ measured by the observer far from the star is
\begin{equation}
E = \delta (1-R_S/R)^{1/2} E'
\end{equation}
which appears similar to equation (\ref{eq:specE}), but includes both the gravitational redshift and the Doppler shift from Equation (\ref{eq:delta}).

{\it Light deflection.}---The relation between $\psi$ and $\alpha$ is \citep{1973grav.book.....M,1983ApJ...274..846P}:
\begin{equation}
\psi(b,R)= \int_{R}^{\infty}\frac{\diff r}{r^2}\left[\frac{1}{b^2}-\frac{1}{r^2}\left(1-\frac{R_S}{r}\right)\right]^{-1/2}
\label{eq:defl}
\end{equation}
where 
\begin{equation}
b=\frac{R}{\sqrt{1-R_S/R}}\sin\alpha
\end{equation}
is the impact parameter of a light-ray originating from the neutron star radius $R$ that is emitted at an angle $\alpha$ with respect to the radial direction, as seen in the local static frame. 

For sufficiently compact stars ($R/R_s\le 1.76$), light ray bending angles can be $>\pi$, resulting in the entire surface being always visible and regions on the neutron star having multiple photon trajectories that can reach the distant observer \citep[see][]{1986ApJ...300..203F}. The code used for the parameter estimation analyses in \cite{miller19} accounts for this possibility. The code used in \cite{riley19} does not do so due to the extra computational complexity required to include multiple images.  We note, however, that the favored ranges of $M$ and $R$ for PSR~J0030$+$0451 from both analyses do not cover the regime of compactness where multiply-imaged regions are relevant.


{\it Time delays.}---Photons with larger deflection angles $\psi$ also travel a larger distance to an observer, which in turn means that their propagation time is longer.  The actual time of propagation to infinity is infinite for any ray, but it is convenient to subtract the propagation time of a radial ray \citep{1983ApJ...274..846P}: 
\begin{equation}
\Delta t(b,R)=\frac{1}{c} \int_{R}^{\infty}\frac{\diff r}{1-R_S/r} \Bigg\{\left[1-\frac{
b^2}{r^2}\left(1-\frac{R_S}{r}\right)\right]^{-1/2}-1\Bigg\} 
\label{eq:delay}
\end{equation}
This time delay translates into a phase lag ($\Delta\phi$) of a photon
\begin{equation}
\Delta \phi =\Omega\Delta t \;,
\end{equation}
and thus the measured rotational phase is $\phi_{\rm obs} = \phi_{\rm emit}+\Delta \phi$ for a photon emitted at phase $\phi_{\rm emit}$
\citep{2004A&A...426..985V}. 

{\it The observed flux.}---When the various factors are combined, the spectral flux from a surface element ${\rm d}S'$ seen by an observer at distance $D$ is (see, e.g., the derivation in section 3.1 of \citealt{2003MNRAS.343.1301P})
\begin{equation}
{\rm d}F(E)=(1-R_S/R)^{1/2}\delta^3 I'(E',\alpha')\cos\alpha'\frac{\diff\cos\alpha}{\diff\cos\psi} \frac{{\rm d}S'}{D^2}\; ,
\label{eq:flux}
\end{equation}
where the surface area element in the co-rotating frame is given by ${\rm d}S' = \gamma R^2 \sin\theta\,{\rm d}\theta\,{\rm d}\phi$, as in the special relativistic case. The lensing factor ${\rm d}(\cos\alpha)/{\rm d}(\cos\psi)$  accounts for the divergence 
of nearby light rays as they propagate away from the star, causing an element of area on the star to appear larger to an observer far from the star. The factor ${\rm d}(\cos\alpha)/{\rm d}(\cos\psi)$ has the limiting value of $(1-R_S/R)$ for light emitted normal to the surface ($\alpha = 0$).  For values of $R_S/R < 0.568$,  as $\alpha$ increases, ${\rm d}(\cos\alpha)/{\rm d}(\cos\psi)$ does as well, but the dependence is non-monotonic for more compact stars.
The number flux ${\rm d}f(E)$ of photons with energy $E$ in the detector is related to the spectral flux by ${\rm d}f(E) = {\rm d}F(E)/E$. 


\subsection{Surface Gravity}

In Schwarzschild geometry, the acceleration due to gravity on the surface of a spherical star is given by the corrected Newtonian formula \citep[see][]{1971reas.book.....Z}
\begin{equation}
g_0=\left(1-\frac{R_S}{R}\right)^{-1/2}\frac{GM}{R^2}.
\label{eq:gsurf}
\end{equation}
The surface gravity is of importance when using realistic emission models such as NS atmospheres, since the equation of state and radiative-transfer properties of the atmosphere are determined in part by this parameter \citep[see, e.g.,][and references therein]{2006ApJ...644.1090H}. As discussed in Section~\ref{sec:OS}, for oblate stars the effective acceleration due to gravity for a given colatitude is expressed relative to $g_0$.

\section{Modeling Hot Spot Emission from Neutron Stars in the Oblate Schwarzschild Approximation}
\label{sec:OS}

For stars that are rotating sufficiently rapidly ($\nu \gtrsim 200$ Hz), 
the rotation-induced oblateness of the stellar surface is significant, and the S+D approximation is inadequate. To address this shortcoming, \citet{2007ApJ...663.1244M} developed the oblate-star Schwarzschild-spacetime (OS) approximation in which the spacetime of the NS is described by the Schwarzschild metric, the special relativistic Doppler boost and aberration and time delays are implemented in the same manner as in the S+D approximation, but the oblateness of the NS surface is also taken into account. In our current OS models that are applied to \textsl{NICER} data, the spin-dependent shape of the rotating star is incorporated through the use of a convenient formula derived by \citet{2014ApJ...791...78A}:
\begin{equation}
R(\theta_c) =  R_{\rm eq} \left[ 1 + o_2(x,\bOmega) \cos^2(\theta_c)  \right],
\end{equation}
where $R_{\rm eq}$ is the equatorial circumferential radius of the neutron star,  $\bOmega = \Omega(2R_{\rm eq}^3/R_S)^{1/2}$, and the expansion coefficient is expressed as $o_{2} = \bOmega^2 ( o_{20} + o_{21} x)$ and is given in Table 1 of \citet{2014ApJ...791...78A}.
\citet{2014ApJ...791...78A}  computed the coefficients $o_n$ using different libraries of proposed EOS and found that the specific choices of tabulated EOS did not significantly alter the values of the coefficients. It should be noted that the choices for the order of the polynomial fits does introduce small differences in the shape function, as can be seen with comparisons with the earlier shape function introduced by
\citet{2007ApJ...663.1244M}. In order to quantify the potential errors introduced by the shape function, pulse shapes using the two different shape functions are compared in the lower panel of Figure \ref{fig:SDvOSvRNS2} with the curve denoted ``Shape Difference". This shows that the potential errors introduced by the shape function are at approximately the 0.1\% level for the test model that spins at 200 Hz. For stars rotating much more rapidly (say at 600 Hz or higher) it would be advisable to investigate an empirical shape function optimized for rapid rotation.

On an oblate star, the surface area of a small spot of angular sizes ${\rm d}\theta_c$ and ${\rm d}\phi$ located at angles $\theta_c$ and $\phi$ on the surface of the star will have a surface area
\begin{equation}
{\rm d}S(\theta_c) = R^2(\theta_c) \sin\theta_c \left[ 1 + f^2(\theta_c) \right]^{1/2} \diff\theta_c \diff\phi
\label{eq:dS}
\end{equation}
where the function $f(\theta_c)$ is defined by
\begin{equation}
f(\theta_c) = \frac{(1-R_S/R)^{-1/2}}{R} \frac{\diff R}{\diff\theta_c}.
\label{eq:f}
\end{equation}
Unlike the spherical case, away from the equator and spin pole, the direction normal to the surface of an oblate star is not generally the radial direction. As a result, it is necessary to consider the photon emission direction $\sigma$ relative to the local zenith angle,
\begin{equation}
\cos\sigma = \sin \alpha \sin \tau \cos \lambda (t) + \cos \alpha \cos \tau\; ,
\end{equation}
where $\tau$ is the angle between the radial direction and the local surface normal, given by
\begin{equation}
\cos \tau =  \left[ 1 + f^2(\theta_c) \right]^{-1/2} \;\textrm{and}\; \sin \tau = f/\sqrt{1+f^2(\theta_c)},
\label{eq:cgamma}
\end{equation}
while spherical trigonometry yields $\cos\lambda(t)=(\cos\zeta-\cos\theta_c\cos\psi)/(\sin\theta_c\sin\psi)$. We note that the angle $\tau$ is positive for points on the ``northern'' hemisphere (i.e., the hemisphere closer to the observer) while $\tau$ is negative for points in the ``southern''
hemisphere.


In the OS approximation, the exterior spacetime remains as the Schwarzschild solution, so the deflection of outward rays can be computed using Equation~\ref{eq:defl}, as in the S+D approximation. Since the surface of an oblate spheroid is tilted with respect to a spherical surface, some outward-directed rays will be eclipsed by the oblate surface.

The oblateness of the star requires special treatment of light rays that originate from the surface in radially-inward trajectories that would be blocked by the stellar surface if the surface were spherical, but can reach the observer due to surface tilting. These photons first travel to smaller values of $r$ until a critical radial coordinate $r_c$ is reached, and then move outwards. The impact parameter of an initially ingoing photon is 
\begin{equation}
b_{\rm in} \le \frac{R(\theta_c)}{\sqrt{1-R_S/R(\theta_c)}} \;,
\end{equation}
for an angle $\alpha > \pi/2$. The critical radius is determined by the solution of the equation
\begin{equation}
r_c =  b_{\rm in} \sqrt{1-\frac{R_S}{r_c}}.
\end{equation}
The bending angle corresponding to a photon trajectory between $R(\theta_c)$ to $r_c$ is given by the integral
\begin{equation}
\Delta\psi = b_{\rm in} \int_{r_c}^{R(\theta_c)} \frac{\diff r}{r^2} 
\left[ 1 - \frac{b_{\rm in}^2}{r^2} \left(1-\frac{R_S}{r}\right)\right]^{-1/2}.
\end{equation}
Owing to symmetry, the inward $R \rightarrow r_c$ and outward $r_c \rightarrow R$ trajectories have the same bending angle, leading to a total bending angle for an initially ingoing photon of
\begin{equation}
\psi_{\rm in}(b_{\rm in},R) = 2 \Delta \psi + \psi(b,R)
\end{equation}
where $\psi(b,R)$ is given by equation (\ref{eq:defl}). 
As expected, an initially ingoing photon will have a larger bending angle and take longer to reach the observer than an initially outgoing photon with the same value of impact parameter. If the star were more compact than $R=3GM/c^2$, some inward-directed photons
would hit the surface instead of escaping to infinity.  However, we
have limited $R/(GM/c^2)$ to be greater than $3.2$, and we find that for the rotation frequencies we explore, no ray that initially moves away from the
surface returns to intersect the surface again.

The flux measured by an observer at a distance $D$ from a surface element on an oblate star is
\begin{equation}
{\rm d}F(E)=(1-R_S/R)^{1/2}\delta^3 I'(E',\sigma')\cos\sigma' \left| \frac{{\partial} \cos\alpha}{{\partial}\cos\psi} \right|_{R} \frac{{\rm d}S'}{D^2}\; ,
\label{eq:flux_os}
\end{equation}
where the area element in the co-rotating frame is, similar to the spherical case, the Lorentz $\gamma$ factor times Equation (\ref{eq:dS}).

\subsection{Photon Propagation Time Delays}

The treatment of travel time delays in the OS approximation is similar to the S+D case, but requires additional corrections. One correction arises from the fact that,  
relative to a photon emitted from the stellar equator, photons at other colatitudes have to travel out towards the observer starting from a smaller radius. Additionally, because photons are emitted from different depths in the gravitational potential of the star (with photons at the spin poles having to climb out further compared to equatorial photons), they will experience different levels of time dilation. Defining a radial photon emitted from the stellar equator as a reference, the total time delays for photons emitted from an oblate star as a function of colatitude can be expressed as:
\begin{equation}
\Delta t(b)=
\frac{1}{c}
\int_{R}^{\infty}
\frac{\diff r}{1-R_S/r}
\Bigg\{
\left[1-\frac{b^2}{r^2}\left(1-\frac{R_S}{r}\right)\right]^{-1/2}-1
\Bigg\}
+
\frac{1}{c}
\left[R_{\rm eq}-R(\theta_c)\right]
+
\frac{1}{c}
R_S\log\left(\frac{R_{\rm eq}-R_S}{R(\theta_c)-R_S}\right) .
\label{eq:osdelay}
\end{equation}
The choice of reference photon for the time delays is arbitrary, but the choice of a fixed location (such as the equator) is appropriate for spots with large angular radius. \citet{2007ApJ...663.1244M} chose a reference photon emitted at the same location as the photon that reaches the observer. However, that choice is only useful for infinitesimal spots.

For initially inward photon trajectories, the time it takes a photon to travel the distance between $R(\theta)$ and $r_c$ is
\begin{equation}
\Delta T =  \int_{r_c}^R \diff r \left(1-\frac{R_S}{r}\right)^{-1}
\left[1 - \frac{b_{\rm in}^2}{r^2}\left(1-\frac{R_S}{r}\right) \right]^{-1/2} \;. 
\end{equation}
By symmetry, when the photon travels from $r_c$ out to $R(\theta)$, the extra time is again $\Delta T$, such that the total time delay for an initially inward ray is
\begin{equation}
T_{\rm in}(b_{\rm in},R) = 2 \Delta T + \Delta t(b_{\rm in},R).
\end{equation}

\subsection{Surface Gravity}

An additional complication of an oblate star is that the effective surface gravity varies with colatitude. \citet{2014ApJ...791...78A} have derived an approximation for the effective acceleration due to gravity on the surface of a rapidly rotating neutron star as a function of colatitude:
\begin{equation}
\frac{g(\theta_c)}{g_0}  =  1 + (c_e \bOmega^2 + d_e \bOmega^4 + f_{e} \bOmega^6) \sin^2\theta_c + 
(c_p \bOmega^2 + d_p \bOmega^4  + f_p \bOmega^6 - d_{60}\bOmega^4) \cos^2\theta_c +  d_{60}\bOmega^4 |\cos\theta_c| \;,
\label{eq:rapidg}
\end{equation}
where $g_0$ is given by Equation~\ref{eq:gsurf} and the values of the coefficients are given in Table 5 of \citet{2014ApJ...791...78A}. (NOTE: using $|\cos\theta_c|$ preserves the symmetry about the spin equator.) This empirical relation provides a good description for neutron stars with a wide range of plausible equations of state. In the analysis of \textsl{NICER} data presented in \cite{miller19} and \cite{riley19}, we use this empirical relation for the surface gravity.

\begin{figure}[t!]
\centering
\includegraphics[clip, trim=2cm 2cm 3cm 2cm,width=0.75\textwidth]{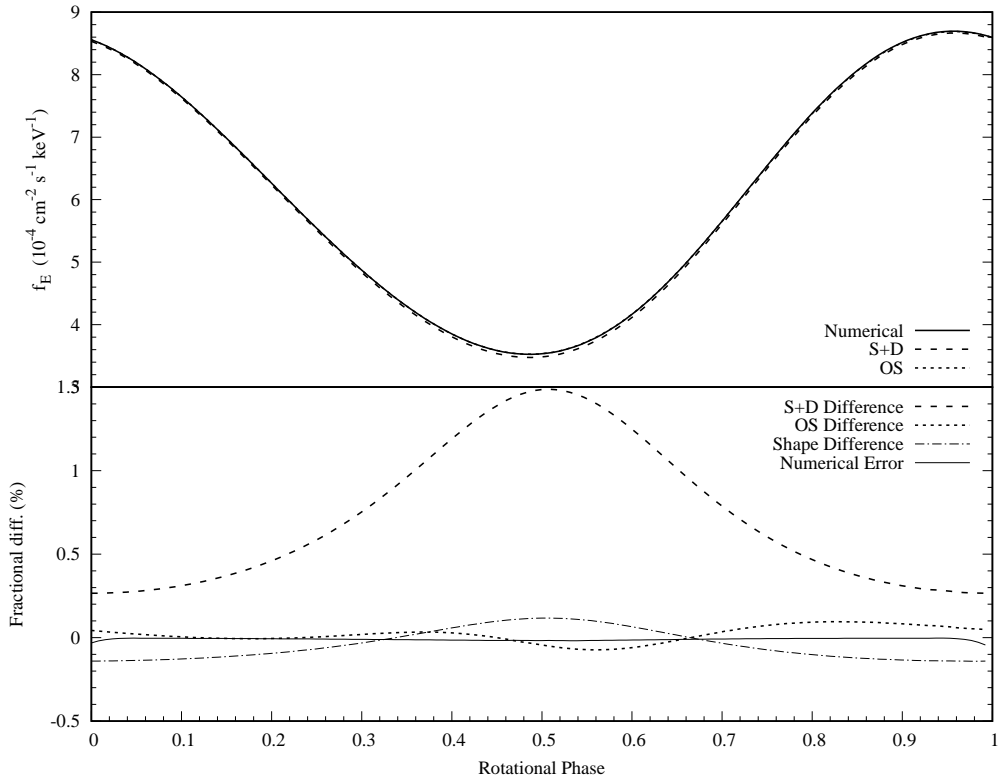}
\caption{Comparison of flux measured at 1~keV for the S+D and OS approximations and the exact numerical waveform for one sample star. The example star spins with a frequency of 200~Hz, has $M=1.44$~\msol, $R_{\rm eq}=11.41$~km. The upper panel shows waveforms computed using the S+D and OS approximations, as well as the exact numerical waveform. The lower panel shows the percent difference between the approximate and exact solutions.
}
\label{fig:SDvOSvRNS2}
\end{figure}

\subsection{Accuracy of the Oblate Schwarzschild Approximation}
\label{sec:accuracy}

We now turn to the accuracy of the OS approximation for NS rotating with spin frequencies similar to the rotation-powered pulsars studied  by \textit{NICER}. This accuracy can be found by computing a pulse waveform with an adaptation of the code described by
\citet{2007ApJ...654..458C}, and comparing with a waveform computed using the OS approximation with the same values of mass, radius, and spin. The code has two main sources of numerical error: one is the spacing of the spatial grid used to store the metric functions, and the other is the geodesic integrator. The geodesics are integrated using a 5th order Runge-Kutta integrator with adaptive step-size.  The magnitude of the numerical error can be estimated by discretizing the non-rotating Schwarzschild metric on the same grid that the rotating spacetime is computed, and using the same tolerances in the R-K integrator. A comparison of the resulting light deflection angles arising from the geodesic integrator with the ``exact" deflection angles computed using equation \ref{eq:defl} yields an indication of the level of error. We chose grid and tolerance levels such that the resulting pulse shapes for Schwarzschild have fractional differences less than 0.01\%. 

The $\sim\,$0.1\% accuracy of pulse waveforms computed using the OS approximation is illustrated in 
Figure \ref{fig:SDvOSvRNS2}. The sample NS's structure is computed using the RNS code \citep{1995ApJ...444..306S} with the \citet{Akmal} equation of state for a spin frequency of 200~Hz, $M=1.44$~\msol and $R_{\rm eq}=11.41$~km.  
The upper panel shows pulse waveforms computed for a blackbody spot with a temperature (in the frame of the star) of 0.35\,keV and an angular radius of 0.01~radians, located at an angle of $60$\deg from the spin axis. The observer is located 200\,pc away at an angle of $30$\deg from the spin axis. The lower panel shows fractional percent differences between the approximations and the numerical computation.
The solid curve corresponds to a full numerical evolution of geodesics using the metric for this star computed using RNS. The dotted curve shows the equivalent OS approximate pulse waveform, while the dashed curve shows the SD approximate waveform where the radius of the spherical star is chosen to be the same as the radial coordinate at the location of the spot on the oblate star ($11.39$~km). Even at this relatively slow rotation rate, the S+D approximation introduces errors at 1.5\% for this example. The OS approximation differs from the numerical computation at the level of about 0.1\%, which mainly comes from inaccuracies from the approximation for the shape of the star. This can be seen from the comparison of waveforms using the OS approximation computed using the \citet{2007ApJ...663.1244M} and \citet{2014ApJ...791...78A} shape functions, which differ at approximately 0.1\%. The numerical error introduced by the geodesic integration is shown as a solid line in the lower panel. The numerical error is estimated by comparing an OS approximation waveform and another waveform using numerical geodesic integrations with the Schwarzschild metric and the same oblate shape
function. The numerical error introduced by the geodesic integrator is about an order of magnitude smaller than the errors introduced by the choice of shape function. 

Comparisons of the OS approximation and exact numerical 
waveforms by \citet{Pihajoki2018} at a much higher rotation rate of 700\,Hz show that the OS approximation introduces larger errors. This is consistent with earlier comparisons by \citet{2007ApJ...654..458C} for stars spinning at 600\,Hz, where the OS approximation introduces errors at the level of a few percent. It may be useful to develop a more accurate approximation for more rapidly rotating NS. However, at this time, none of the rotation-powered X-ray pulsars considered as \textit{NICER} targets for $M$-$R$ and dense matter equation of state parameter estimation analysis rotate this rapidly. 

Comparisons of our OS pulse waveforms with exact numerical waveforms show that the OS waveforms agree with the corresponding exact numerical waveforms to better than 0.1\% for spin frequencies $\lesssim 300$~Hz. This accuracy is more than adequate for the purposes of the \textsl{NICER} mission. (All the pulsars currently being considered for \textsl{NICER} pulse waveform analyses have spin frequencies $\lesssim 300$~Hz. A change in the assumed distance by 0.1\% would eliminate much of the remaining difference between the OS waveforms and the corresponding exact numerical waveforms.) The excellent agreement of waveforms computed using the OS approximation with the corresponding exact numerical waveforms shows that the effects on the waveform of frame dragging and the stellar mass quadrupole, which are not included in the OS approximation, are negligible for our purposes. Based on this, for the analysis of \textsl{NICER} targets to constrain the neutron star mass-radius relation, we consider the OS approximation.

\section{S+D Code Verification Tests}
\label{sec:verification}

As noted previously, an important prerequisite for obtaining reliable constraints of the NS mass-radius and dense matter equation of state relation via pulse waveform fitting is that the codes used for this purpose produce both accurate and precise results. We have therefore devised a suite of tests to evaluate different components of the model to which we subject our codes. These include the representation of the hot spot on the stellar surface (both for point-like and extended spots), the general and special relativistic effects, and the calculation of the observed phase-dependent flux. The codes used in these comparisons have been independently developed by several groups: the Columbia University (CU) code, the Illinois-Maryland (IM) code, the two NASA Goddard Space Flight Center (GSFC-S and GSFC-M) codes, and the University of Alberta (AB) code. For the OS analysis described in Section~\ref{sec:ostests}, results from the University of Amsterdam (AMS) suite of codes are also included. All codes are described in Appendix~\ref{sec:codes}.

For consistency, in all codes used in the verification tests we use the up-to-date published values for the various physical and astrophysical constants (e.g., Planck constant, Boltzmann constant,  parsec, $G$\msol) from the Particle Data Group handbook.\footnote{\url{http://pdg.lbl.gov/2017/reviews/rpp2017-rev-phys-constants.pdf}}\textsuperscript{,}\footnote{\url{http://pdg.lbl.gov/2017/reviews/rpp2017-rev-astrophysical-constants.pdf}}
It is important to note that the quantity $G$\msol is determined to much higher precision than $G$ and \msol individually so in all instances the use of the product $G$\msol is recommended.

To evaluate the performance of the codes under consideration, we use two metrics: i) the fractional difference of the output photon flux at each spin phase from the reference code and each other code and ii) the difference between the flux for each phase from the reference code and a given code, divided by the median flux over all phases from the reference code. For these tests, we have chosen a target fractional precision of 0.1\% for both metrics.  We selected this number because \textsl{NICER} observations of individual sources will collect up to millions of counts, and thus the average number of counts over the hundreds of phase-energy bins that we use will be in the thousands, with some bins having tens of thousand counts.  Thus, Poisson fluctuations will be a few tenths of a percent per bin, and a pulse waveform precision of better than 0.1\% guarantees that waveform inaccuracies will not dominate the uncertainties.  Such precision also means that code inaccuracies will be small compared to the desired $\sim$5\% mass-radius measurement precision with \textsl{NICER}. We now describe each test and summarize the outcome of the code validation exercises.

In all cases, we consider a NS with $M=1.4$~\msol and $R=12$ km at a distance of $D=200$ pc, and a surface hot spot at a temperature $kT=0.35$ keV as measured in the surface comoving frame. We define rotational phase $\phi=0$ as the closest approach of the hot spot to the observer.

\begin{figure}[t!]
\begin{center}
  \includegraphics[clip, trim=0cm 6.5cm 0cm 2cm,angle=0,width=0.47\textwidth]{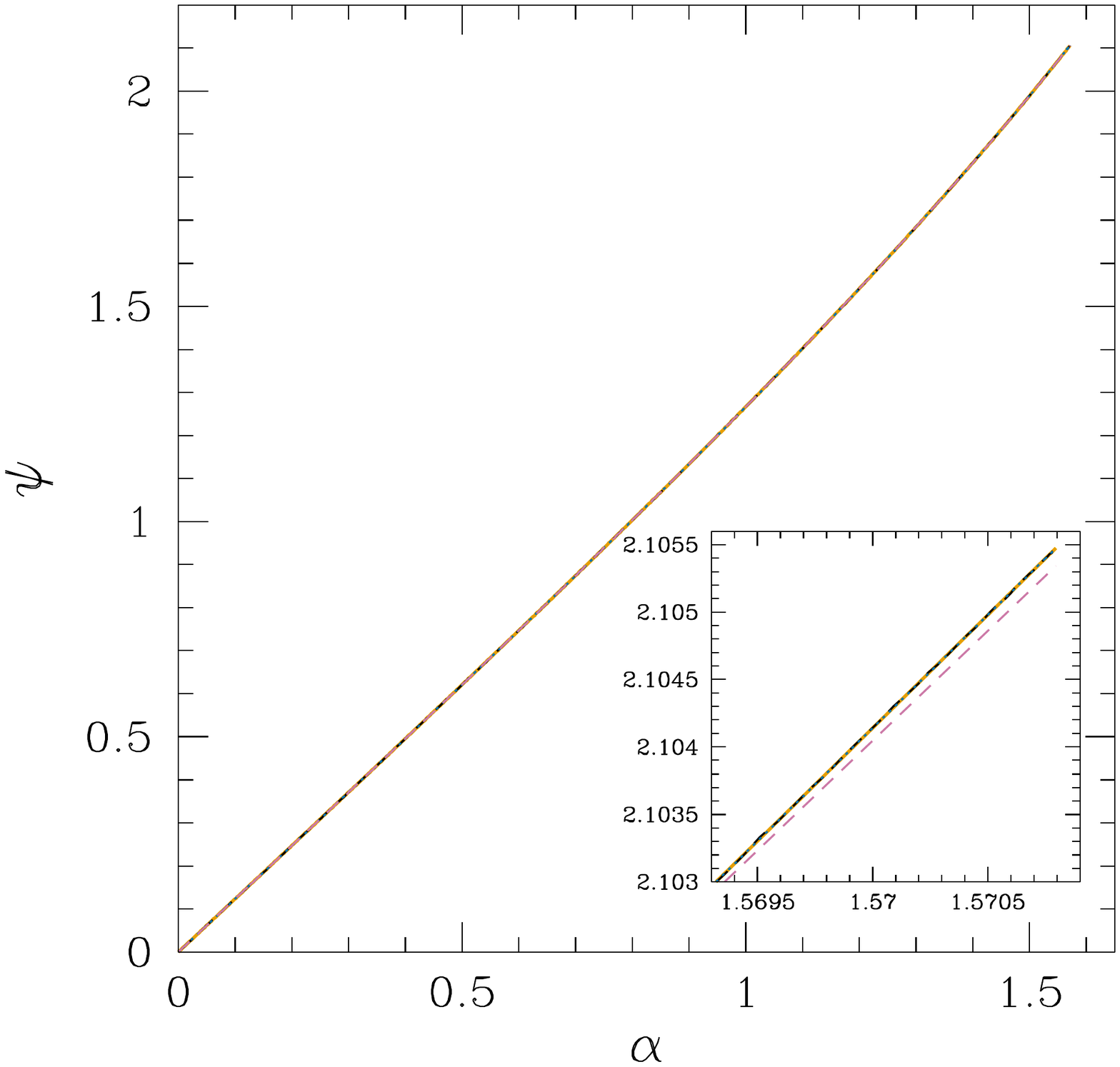}
\includegraphics[clip, trim=0cm 6.5cm 0cm 2cm, angle=0,width=0.47\textwidth]{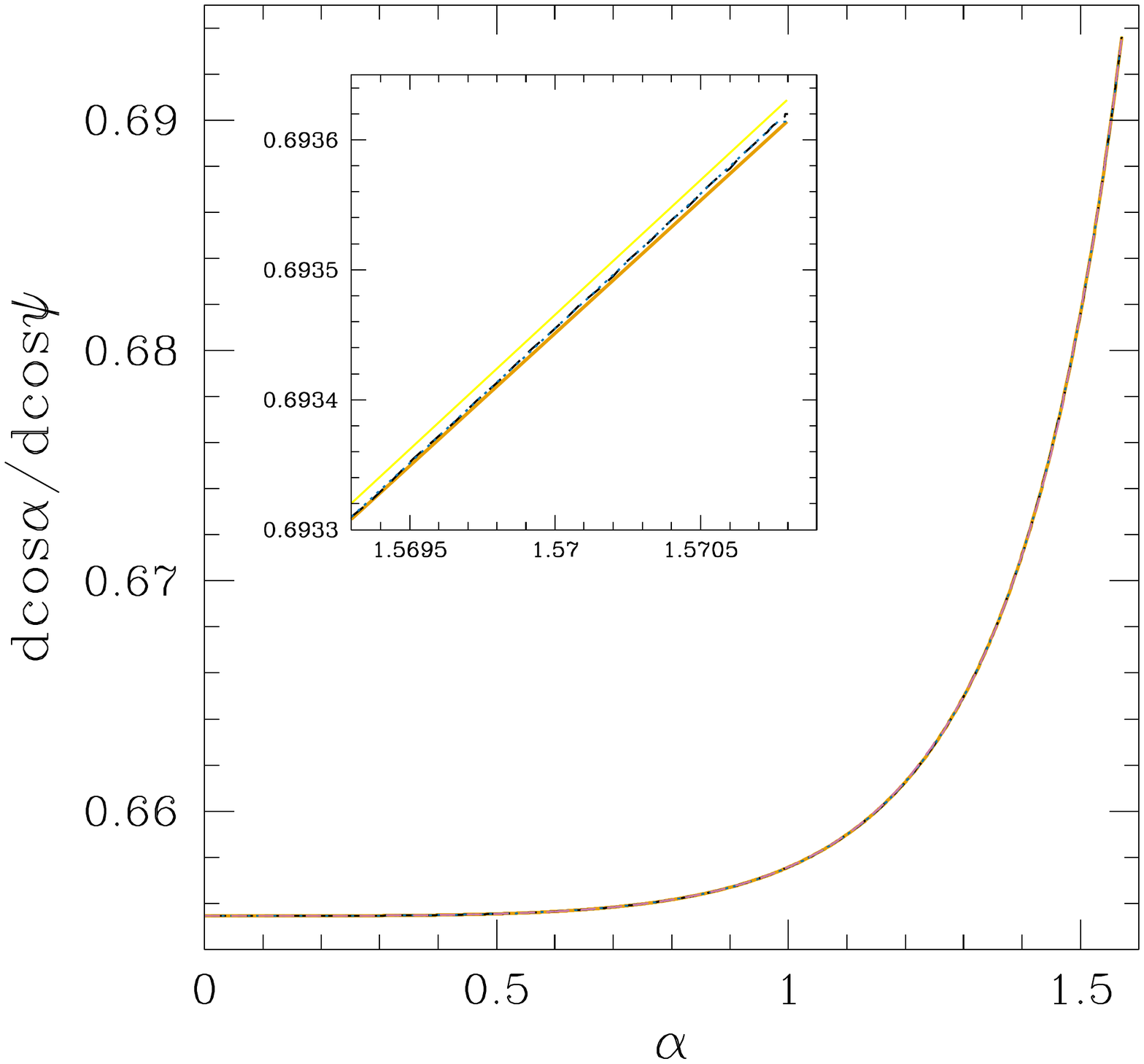}
\caption{Comparison of the calculated light ray deflection angle from Equation~(\ref{eq:defl}) (left) and the lensing factor $d\cos\alpha /d\cos\psi$ that appears in Equation~(\ref{eq:flux}) (right) as a function of the light ray emission angle $\alpha$. The insets show a zoom-in around $\alpha=\pi/2$, where the largest discrepancies between the codes are apparent. The line colors correspond to the results generated by the CU (black), GSFC-M (orange), GSFC-S (blue), Alberta (purple), and IM (yellow) codes.  The different curves are mostly indistinguishable because the agreement between the codes is excellent. 
} 
\label{fig:ray}
\end{center}
\end{figure}

\begin{figure}[t!]
\begin{center}
  \includegraphics[clip, trim=0cm 7cm 0cm 2cm, angle=0,width=0.49\textwidth]{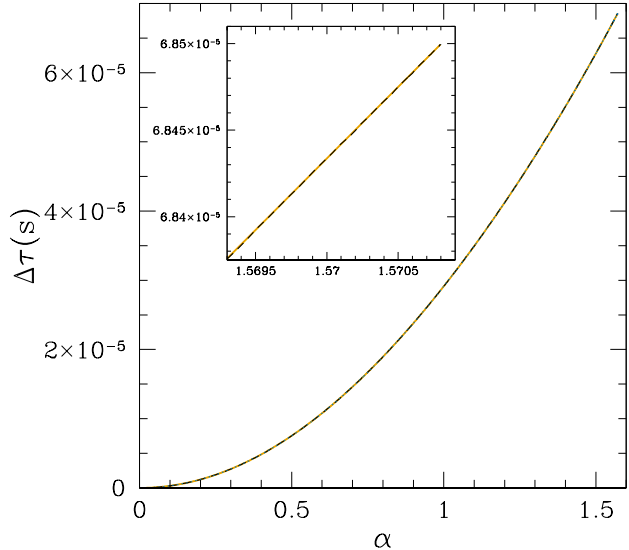}
\caption{Comparison of the calculations of the travel time delay integral from Equation~(\ref{eq:delay}) using the different codes considered.  The inset shows a zoom-in around $\alpha=\pi/2$, where the largest discrepancies between the codes are expected. The color code is the same as in Figure~\ref{fig:ray}. Again, the agreement between the codes is superb.}
\label{fig:delay}
\end{center}
\end{figure}

\subsection{Comparison of Light Deflection, Lensing, and Travel Time Delay Results}

The first comparison involves three key ingredients of the model: the deflection angle given by Equation~(\ref{eq:defl}), the ``lensing factor'' ${\rm d}\cos\alpha/{\rm d}\cos\psi$ from Equation~(\ref{eq:flux}), and the travel time delay difference as a function of emitted angle relative to a radial photon, which is given by Equation~(\ref{eq:delay}). 
A value of $GM/(Rc^2)=0.1723$ was used for all of these plots.
As is clear from Figures~\ref{fig:ray} and \ref{fig:delay}, the agreement between the outputs is excellent. For the deflection angle computation, the difference between the outputs is $\lesssim$0.0001\%, and for the lensing factor it is $\lesssim$0.001\%. The largest discrepancies for the time delay are at a level of $\lesssim$0.0001\%. Therefore, for all practical purposes, the light deflection, time delay, and lensing factors are identical between the codes.

\subsection{Pulse Waveform Test SD1: Rotating Neutron Star with Planck Spectra and Isotropic Emission}

The parameters of our first set of Schwarzschild+Doppler pulse waveform comparisons are summarized in Table~\ref{table:test1}.  We designate the tests SD1a through SD1f, where ``SD'' indicates that the test is for S+D waveforms.  Collectively, these comparisons test the following aspects of the pulse waveform generation codes:

\begin{itemize}

\item The treatment of special relativistic effects such as redshifts/blueshifts and aberration.  The tests include rotational frequencies of 1, 200, and 400~Hz as seen at infinity, which is broader than the range of frequencies for the best candidate \textsl{NICER} sources. Although, as shown in Section~\ref{sec:accuracy}, the accuracy of the S+D approximation becomes poor at $\gtrsim 200$ Hz, for the purposes of these code comparisons, considering faster spins offers enhanced sensitivity to any discrepancies in the implementation of special relativistic effects.

\item The incorporation of general relativistic effects such as light deflection, time delays, and lensing.

\item The treatment of occultations and the use of the full range of photon emission angles (which is the motivation behind choosing $\zeta=\theta_c=90$\deg for four of the tests).

\item The capability of the codes to handle both small and large spots, over the entire range of plausible spot sizes.

\item The ability of the codes to compute pulse waveforms for more general configurations of the spot and observer, including a case in which the hot spot encompasses the rotational pole (Test~SD1f).

\end{itemize}

In tests SD1a-f we compute a monochromatic pulse profile in units of photons cm$^{-2}$ s$^{-1}$ keV$^{-1}$ at 1 keV as measured in the rest frame of the observer. We note that we use a blackbody spectrum and isotropic beaming in these tests.  The true spectrum will differ somewhat in shape, and significantly in normalization, from a blackbody spectrum, and isotropic beaming is not expected in any realistic circumstance.  We nonetheless used a blackbody spectrum and isotropic beaming because the simplicity of these assumptions means that we can perform high-precision analytic checks to the answers when the rotation is slow (1~Hz) or at special phases and geometries (e.g., when the spot is small and is directly under an equatorial observer).  This allows us to check the normalizations of the outputs of the codes, as well as other results related to the sharp line tests (see Section~\ref{sec:lines}).
The results from all codes agree with the analytic expectations to high precision as described in Appendix~\ref{app:SDanalytical}. Some of these tests, as well as others not listed here, are also discussed in Appendix~A of \citet{2013ApJ...776...19L}.


\begin{table}[h]
\begin{center}
\footnotesize{
  \begin{tabular}{lcccccc}
    \hline
    \emph{Quantity} & \emph{Test SD1a} & \emph{Test SD1b} & \emph{Test SD1c} & \emph{Test SD1d} & \emph{Test SD1e} & \emph{Test SD1f}\\
    \hline
    Number of hot spots              & 1   & 1   & 1    & 1 & 1 & 1\\
    Colatitude of spot center (\degr) & 90  & 90  & 90   & 90 & 60 & 20\\
    Angular radius of hot spot (rad)  & 0.01 & 1.0 & 0.01 & 1.0 & 1.0 & 1.0\\
    Colatitude of observer (\degr) & 90  & 90  & 90   & 90 & 30 & 80\\
    Neutron star mass (\msol)     & 1.4 & 1.4 & 1.4  & 1.4 & 1.4 & 1.4\\
    Neutron star radius (km)         & 12 & 12   & 12   & 12 & 12 & 12\\
    $\nu$ at infinity (Hz)           &  1 &  1   & 200  & 200 & 400 & 400\\
    Spectrum of emission             & Planck & Planck & Planck & Planck & Planck & Planck \\
    Beaming of emission              & iso & iso & iso & iso & iso & iso\\
    Temperature of emission (keV)    & 0.35 & 0.35   & 0.35 & 0.35 & 0.35 & 0.35\\
    \hline
  \end{tabular}
  \caption{Parameter values for Waveform Test SD1.
}
    \label{table:test1}}
\end{center}  
\end{table}

\begin{figure}[t!]
\begin{center}
  \includegraphics[clip, trim=5.5cm 5.5cm 1cm 3cm,angle=0,width=0.42\textwidth]{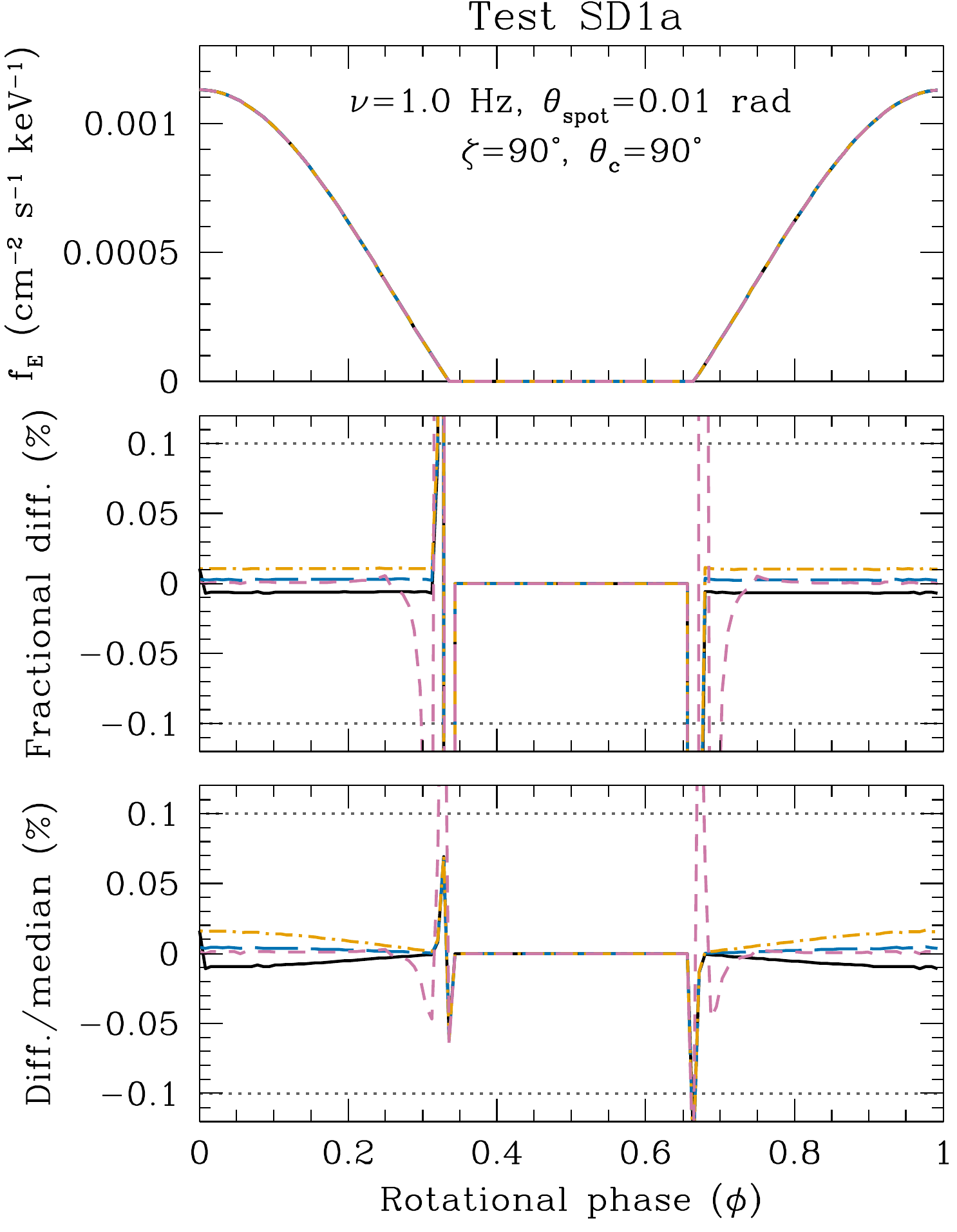}~~~
  \includegraphics[clip, trim=5.5cm 5.5cm 1cm 3cm,angle=0,width=0.42\textwidth]{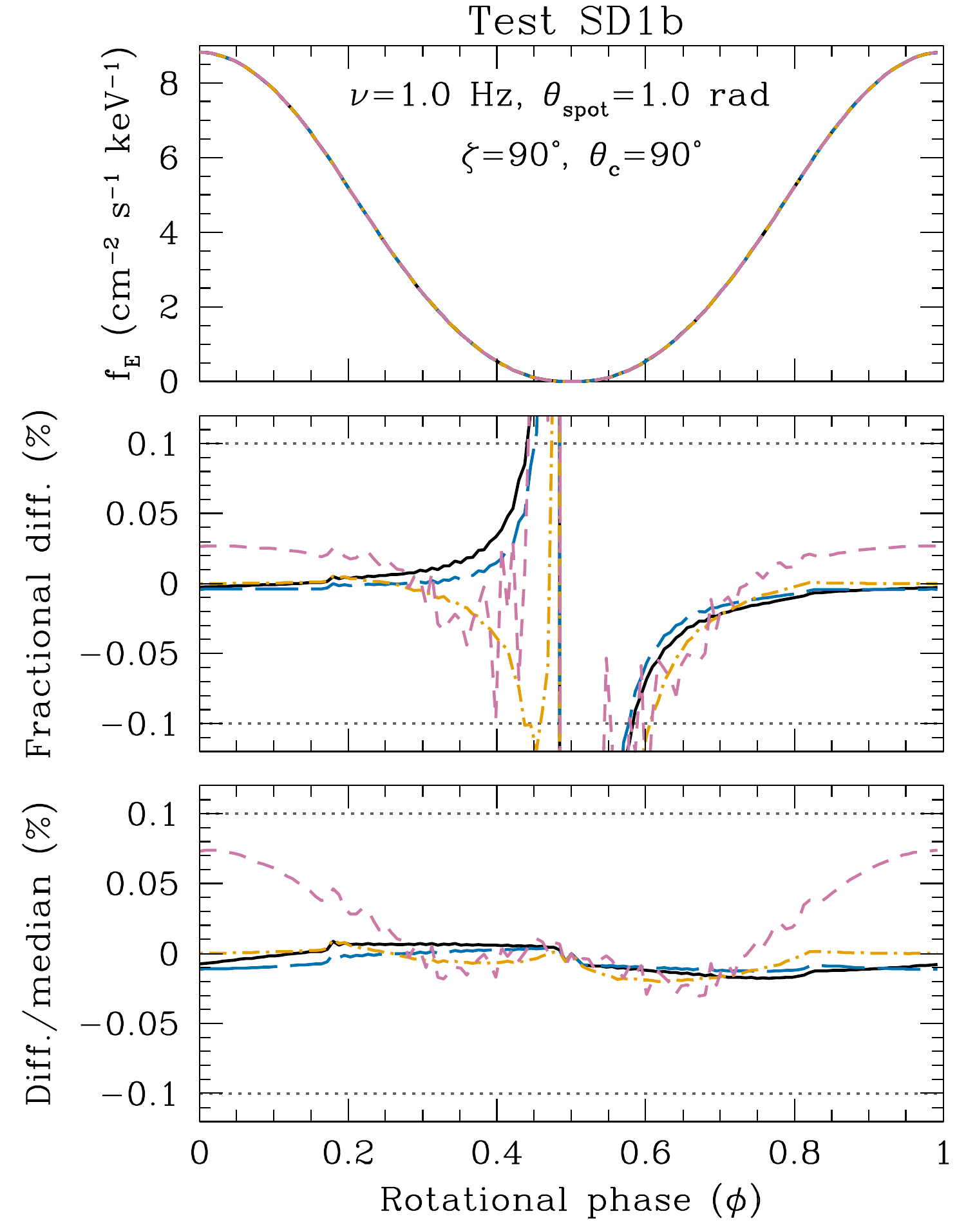}
  \caption{Comparisons of the synthetic pulse waveforms from Tests~SD1a and SD1b, the slowly spinning (1 Hz) variations of the SD1 tests (see Table~\ref{table:test1} for the assumed parameters). The point-like ($\theta_{\rm spot}=0.01$ rad) and large spot ($\theta_{\rm spot}=1$ rad) are shown on the left and right, respectively. The top panel shows the pulse waveforms. The middle panel shows the fractional difference between the CU (black), GSFC-M (orange), GSFC-S (blue), and Alberta (purple)  photon fluxes compared to the IM flux at each phase bin, expressed as a percentage. In the bottom panel, the IM flux is subtracted from the other fluxes and the result is divided by the median IM flux over all phases. The two horizontal dotted lines mark the target $\pm$0.1\% measurement precision.  Except near the spot eclipse ingress and egress, where the flux is two orders of magnitude smaller than it is at the peak, the agreement between the codes is significantly better than the target precision. 
  }
\label{fig:test12}
\end{center}
\end{figure}

\begin{figure}[t!]
\begin{center}
  \includegraphics[clip, trim=5.5cm 5.5cm 1cm 3cm,angle=0,width=0.42\textwidth]{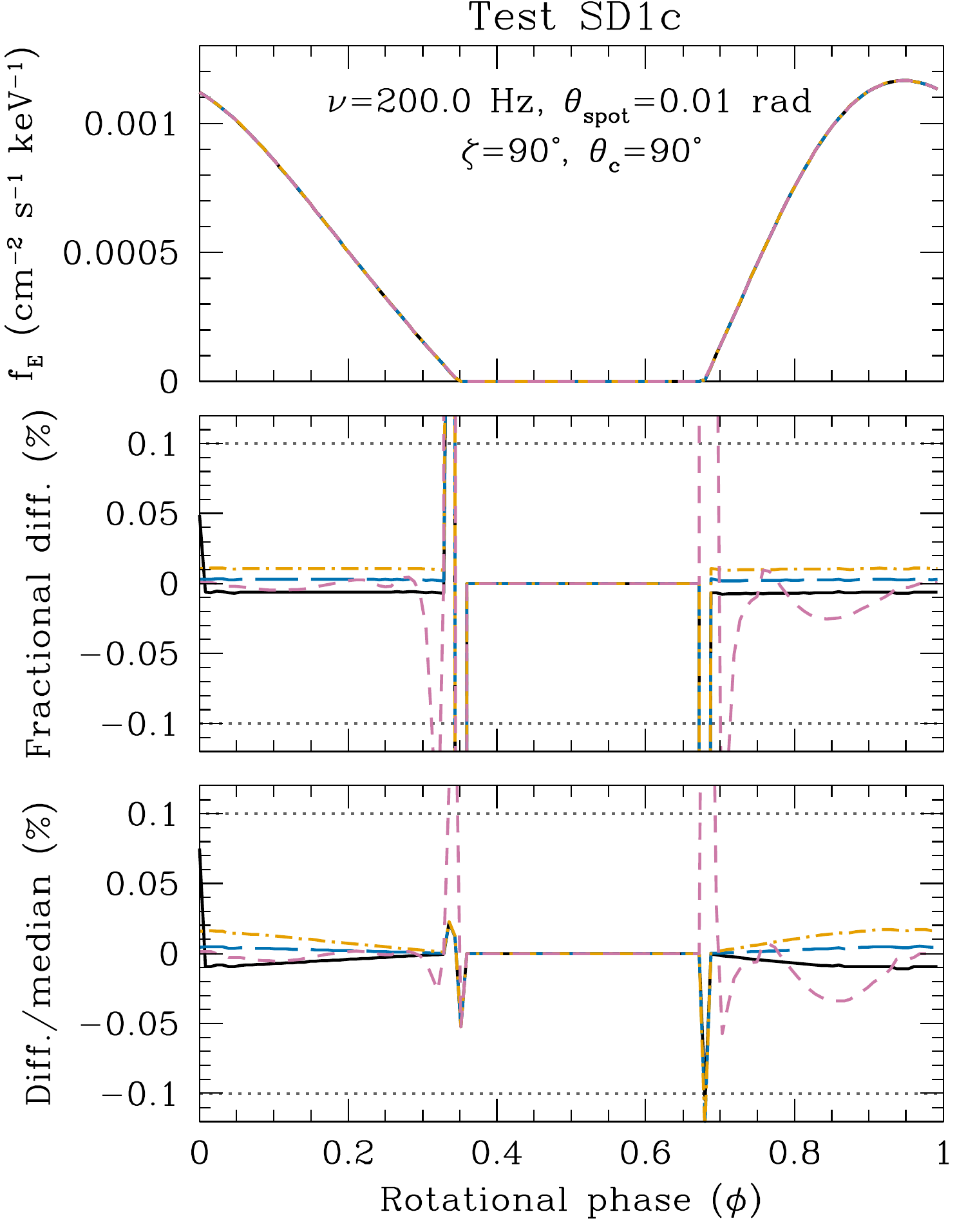}~~~
  \includegraphics[clip, trim=5.5cm 5.5cm 1cm 3cm,angle=0,width=0.42\textwidth]{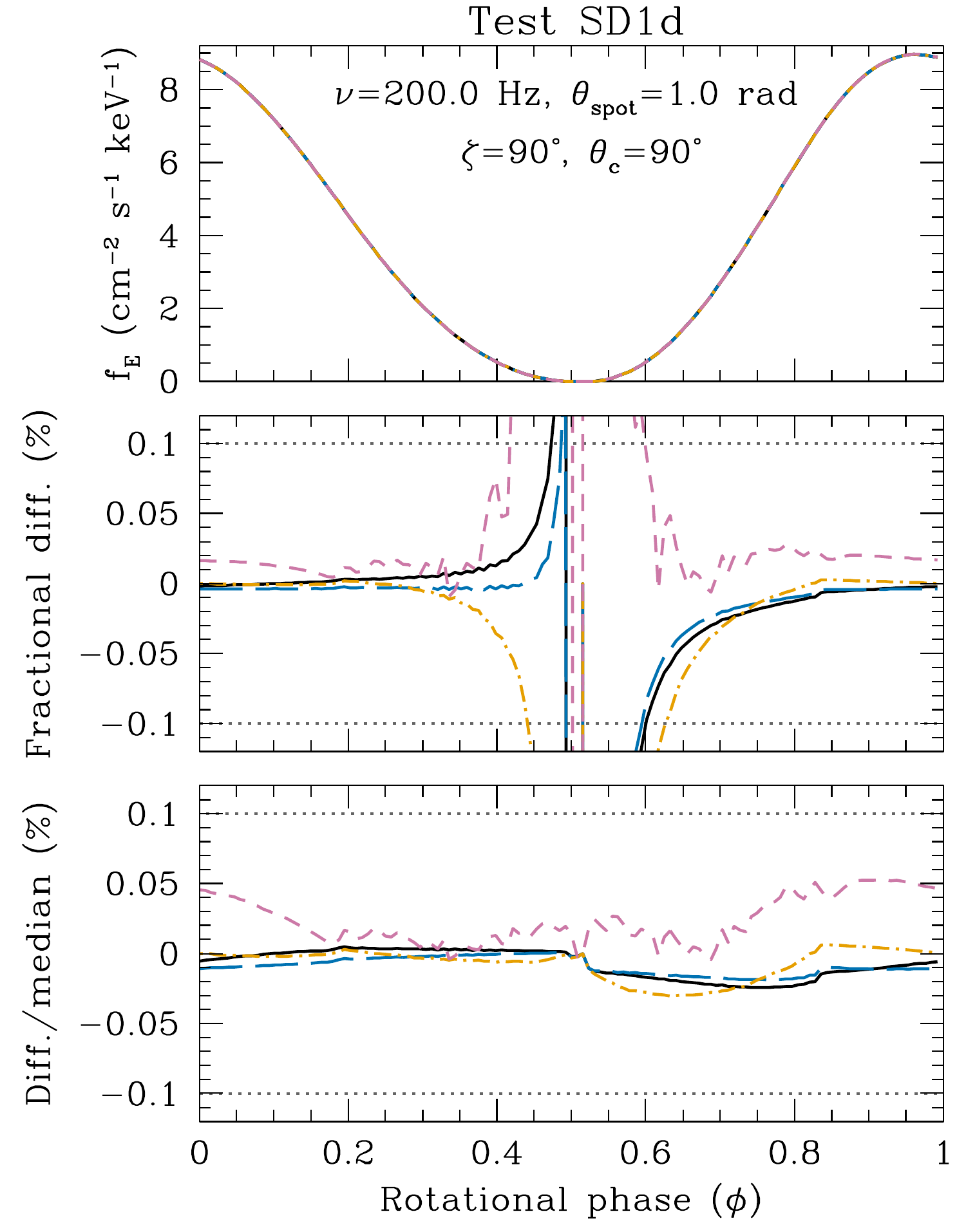}
  \caption{Same as Figure~\ref{fig:test12} except for tests SD1c and SD1d, which consider more rapid rotation ($\nu = 200$~Hz).}
\label{fig:test34}
\end{center}
\end{figure}

\begin{figure}[t!]
\begin{center}
  \includegraphics[clip, trim=5.5cm 5.5cm 1cm 3cm,angle=0,width=0.42\textwidth]{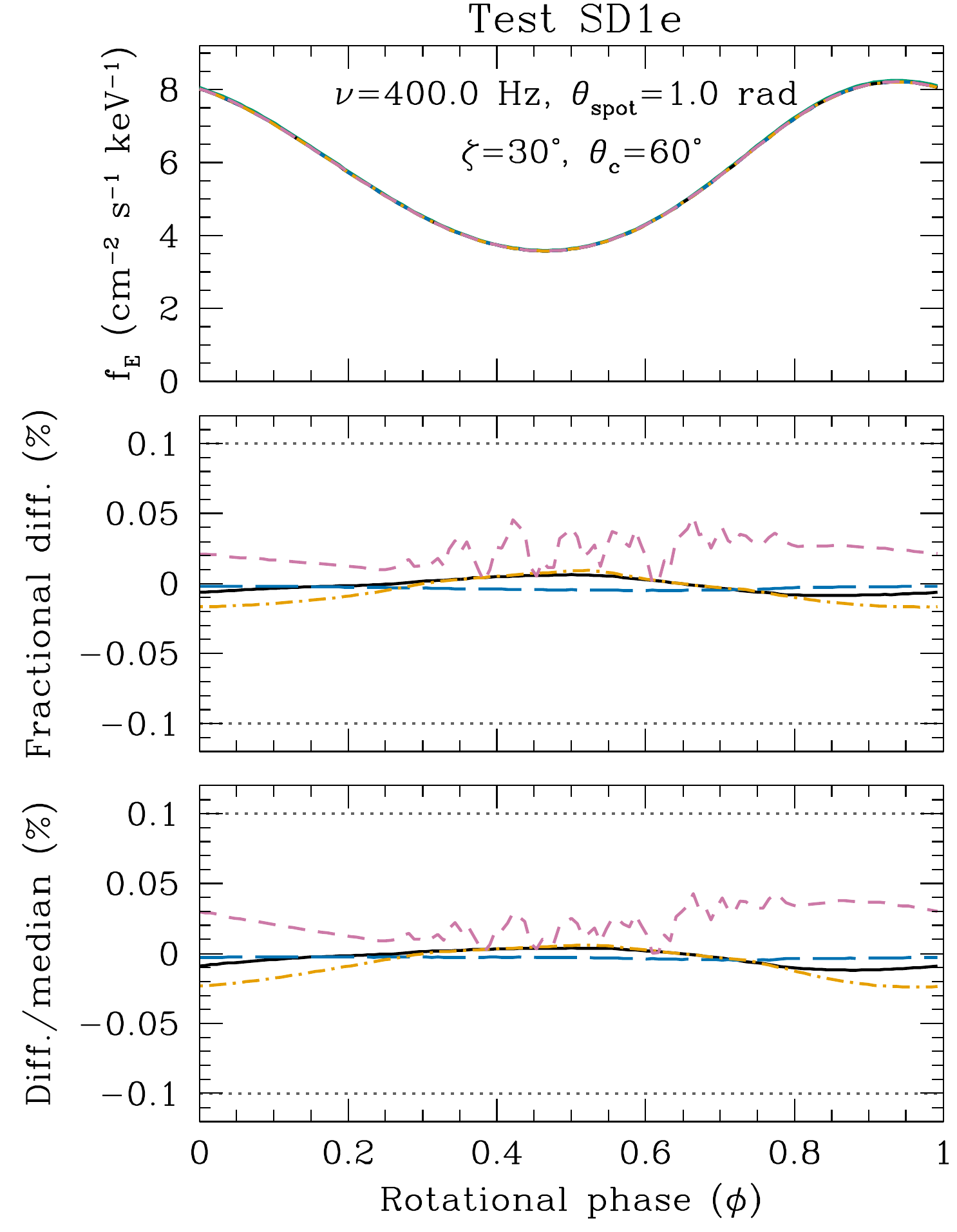}~~~
  \includegraphics[clip, trim=5.5cm 5.5cm 1cm 3cm,angle=0,width=0.42\textwidth]{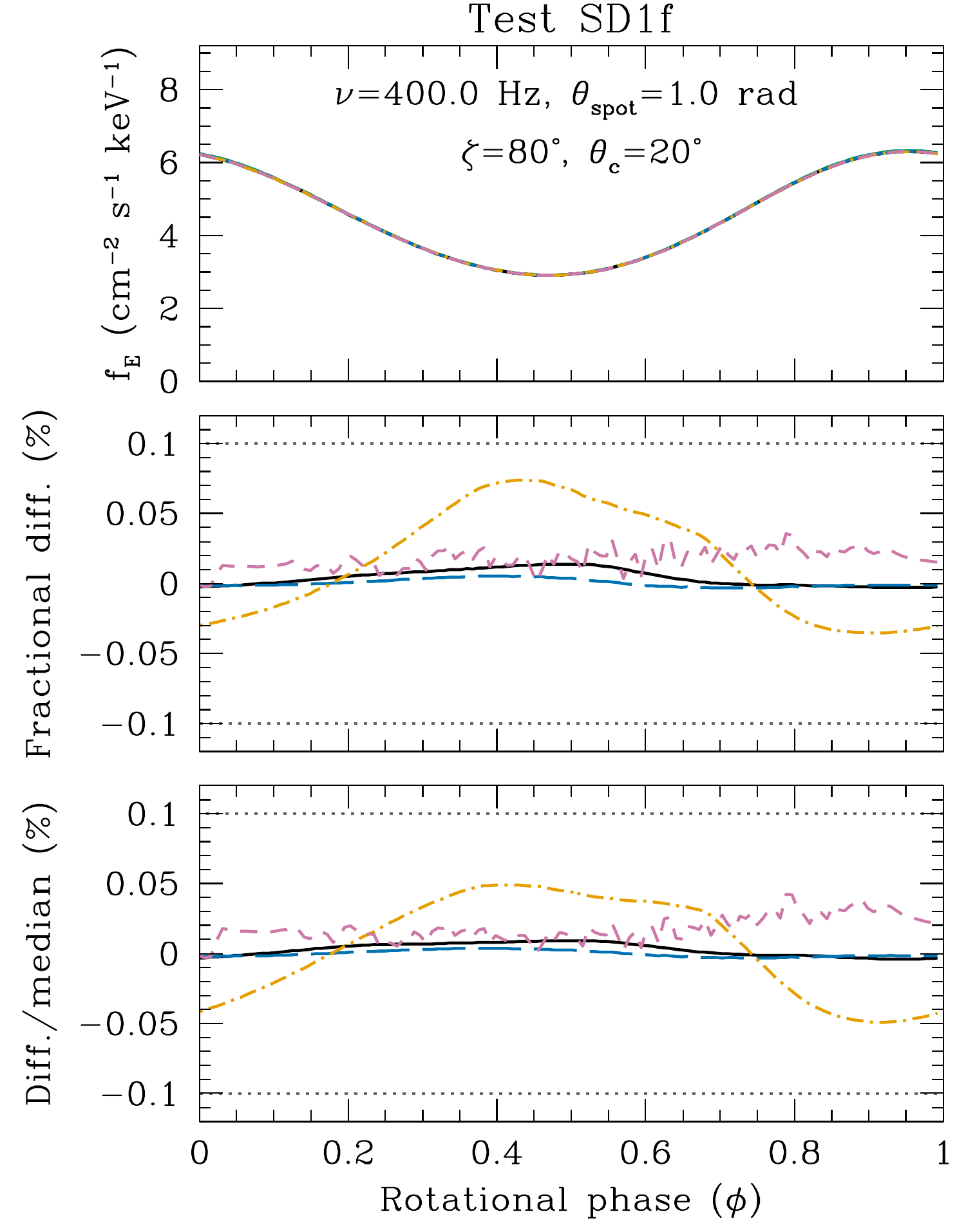}
  \caption{Same as Figure~\ref{fig:test12} but for comparisons of the synthetic pulse waveforms from Tests~SD1e and SD1f, which consider a rapidly spinning (400 Hz) neutron star and more general combinations of co-latitude and viewing angle.}
\label{fig:test56}
\end{center}
\end{figure}

Figures~\ref{fig:test12}, \ref{fig:test34}, and \ref{fig:test56} show the results of Tests~SD1a-f.  In all instances, the IM code was used as a reference against which the other codes were compared. It is apparent that all codes perform extremely well in these tests---for most phases, the flux discrepancies are well within the 0.1\% requirement (indicated by the pair of horizontal dotted lines). The most pronounced differences (reaching up to $\sim1\%$) occur near the flux minimum around the ingress and egress as the hot spot is eclipsed by the neutron star. However, in these cases the fluxes are more than two orders of magnitude smaller than the flux around the pulse maximum.  These discrepancies are therefore unimportant for practical purposes.  Overall, the tested waveforms agree with each other, and with analytical results where appropriate, to significantly better than the 0.1\% target precision.

\subsection{Waveform Test~SD2: Rotating Neutron Stars with Isotropic Narrow Line Emission}
\label{sec:lines}

For our second set of comparisons we computed phase-energy waveforms assuming that the emission spectrum is confined to narrow lines.  This enables tests of the following aspects of the codes:

\begin{itemize}

\item Gravitational and special relativistic redshifts.  Because the photon energy at emission is known precisely, it is possible to compute analytically the received energy as a function of the phase of emission and the rotational frequency of the star.

\item Light deflection and time delays.  Using a sharp line, especially from a small spot, it is possible to compute from tables the observed phases spanned by the eclipse.  This provides an additional test of the computation of light bending and time delays.

\item Gravitational lensing.  The intensity observed at a given phase also depends on the lensing factor ${\rm d}\cos\alpha/{\rm d}\cos\psi$.  This can be compared with the computed intensity.

\item The capacity of the codes to deal with lines with sharp energy profiles.  No surface atomic lines have yet been confirmed from any neutron star, but if one were seen then it would provide valuable information.  Real line profiles will not be sharp, but tests involving sharp lines stress the codes as much as possible.

\end{itemize}

We tabulate the model parameters in Table~\ref{table:test2}.


\begin{table}[h]
\begin{center}
  \begin{tabular}{lcccc}
    \hline
    \emph{Quantity} & \emph{Test SD2a} & \emph{Test SD2b} & \emph{Test SD2c} & \emph{Test SD2d} \\
    \hline
    Number of hot spots              & 1   & 1   & 1    & 1\\
    Colatitude of spot center (\degr) & 90  & 90  & 90   & 90\\
    Angular radius of hot spot (rad)  & 0.01 & 1.0 & 0.01 & 1.0\\
    Colatitude of observer (\degr) & 90  & 90  & 90   & 90\\
    Neutron star mass (\msol)     & 1.4 & 1.4 & 1.4  & 1.4\\
    Neutron star radius (km)         & 12 & 12   & 12   & 12\\
    $\nu$ at infinity (Hz)           &  1 &  1   & 400  & 400\\
     Spectrum of emission             & line Planck & line Planck & line Planck & line Planck \\
    Temperature of emission (keV)    & 0.35 & 0.35   & 0.35 & 0.35\\
    Energy range of line (keV)  & 0.99998-1.00002 & 0.99998-1.00002 & 0.995-1.005 & 0.995-1.005\\
    \hline
  \end{tabular}
  \caption{Parameters of Waveform Test SD2.
  \label{table:test2}}
\end{center}  
\end{table}

We assume that the emission line has the specific intensity of a
Planck spectrum with $kT_{\rm eff}=0.35$~keV (as measured in the
surface comoving frame) within a very narrow energy band, and zero
flux outside of that band. For the 1~Hz tests the energy band was
$0.99998-1.00002$~keV as measured in the surface comoving frame, the
model output was for 200 photon energies, as seen by a distant
observer, uniformly spaced in the range 0.809$-$0.81099~keV, and the
pulse waveforms were split in 128 equally spaced phase segments. For
the 400~Hz tests the energy band was 0.995$-$1.005~keV as measured in
the surface comoving frame and the simulations were performed for 200
photon energies, as seen by a distant observer, that were uniformly
spaced in the range 0.7$-$1.0~keV. In all cases comparisons were
performed between the calculated photon fluxes, which were reported in
units of photons~cm$^{-2}$~s$^{-1}$~keV$^{-1}$.

\begin{figure}[t!]
\centering
  \includegraphics[clip, trim=3cm 19.2cm 1cm 3.7cm,angle=0,width=0.495\textwidth]{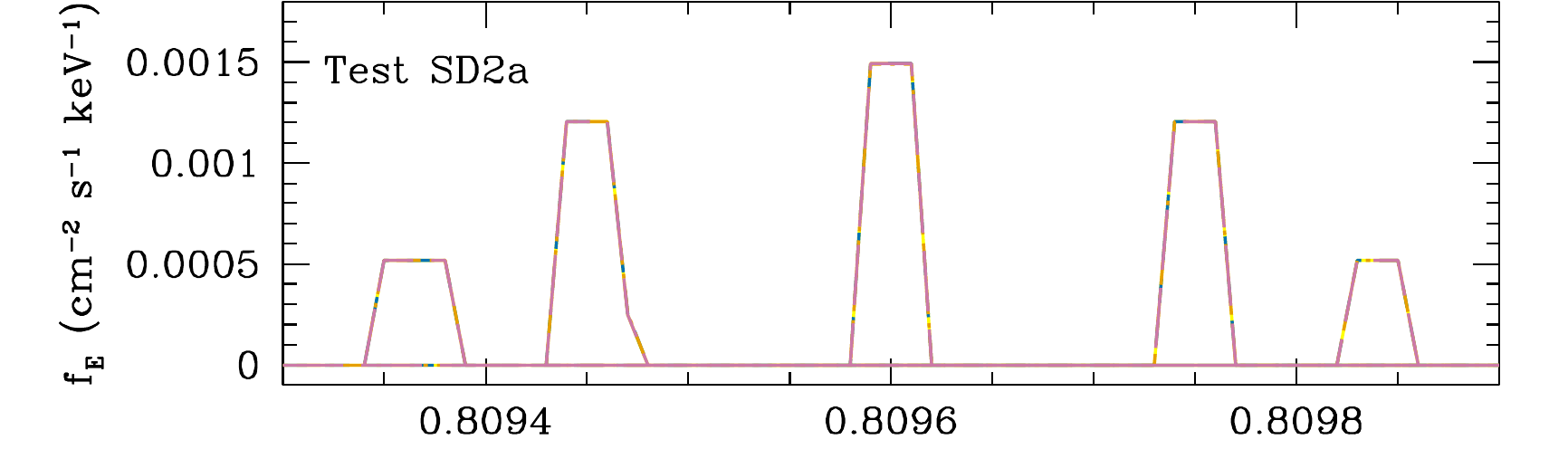}
  \includegraphics[clip, trim=4.2cm 19.2cm 1cm 3.7cm,angle=0,width=0.46\textwidth]{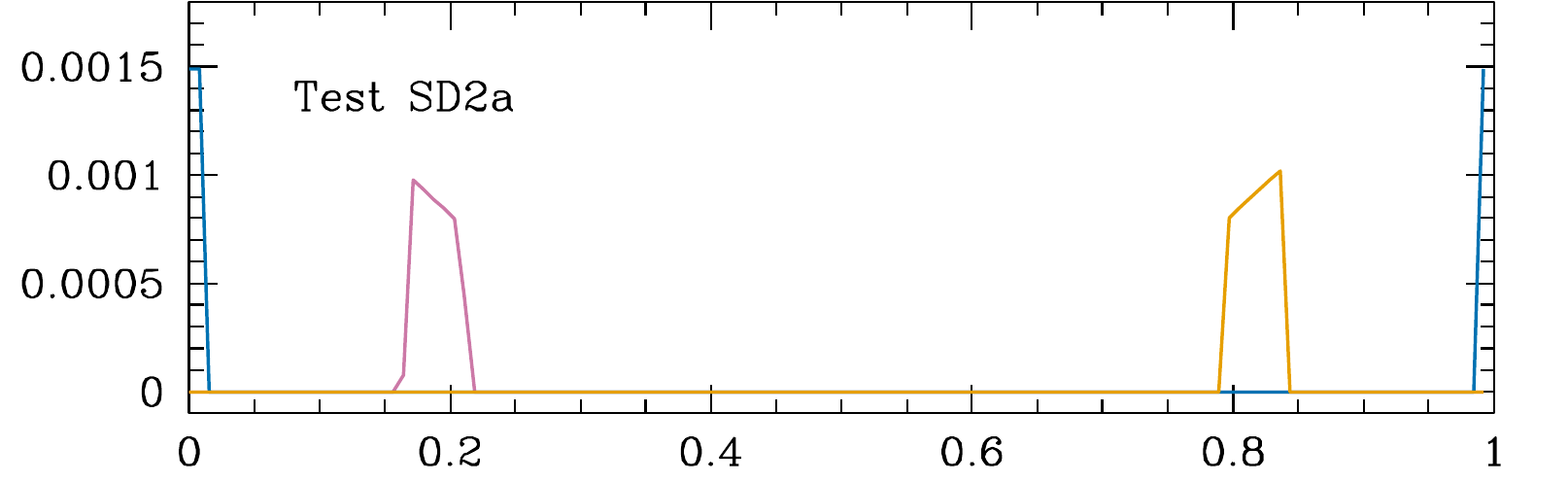}
  \includegraphics[clip, trim=3cm 19.2cm 1cm 3.7cm,angle=0,width=0.495\textwidth]{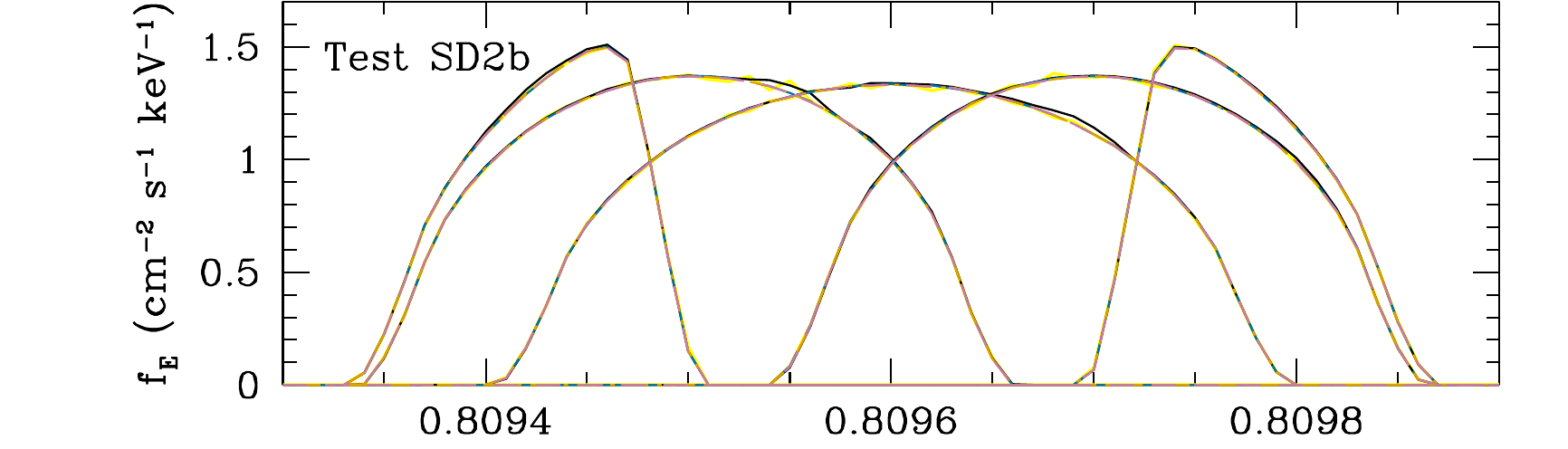}
   \includegraphics[clip, trim=4.2cm 19.2cm 1cm 3.7cm,angle=0,width=0.46\textwidth]{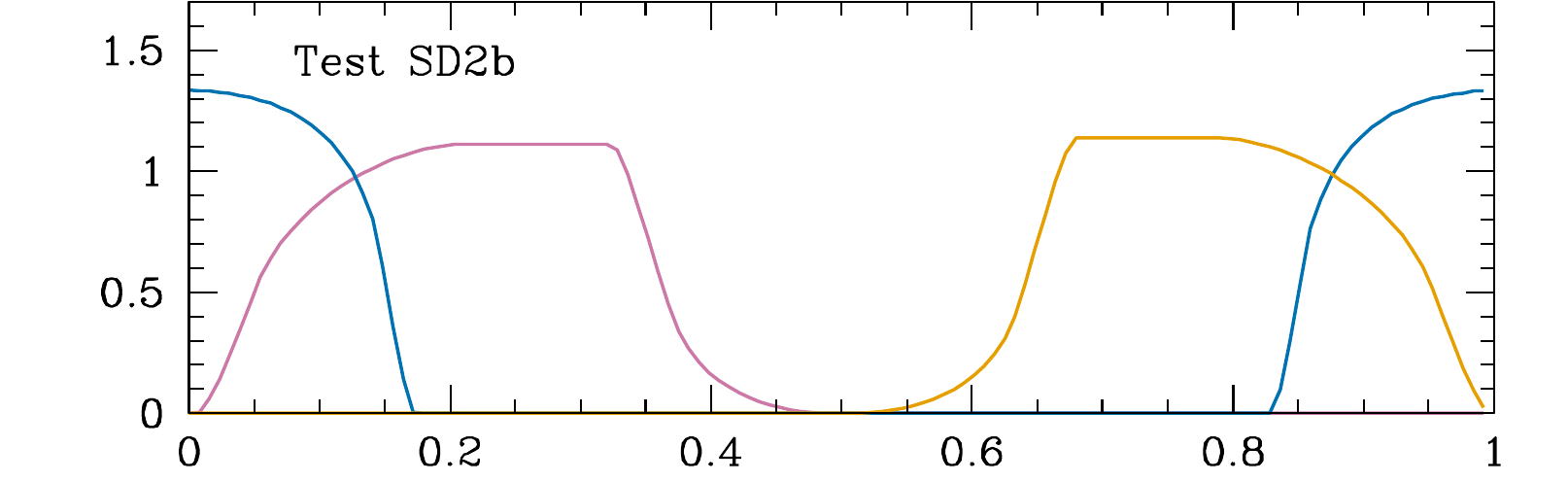}
  \includegraphics[clip, trim=3cm 19.2cm 1cm 3.7cm,angle=0,width=0.495\textwidth]{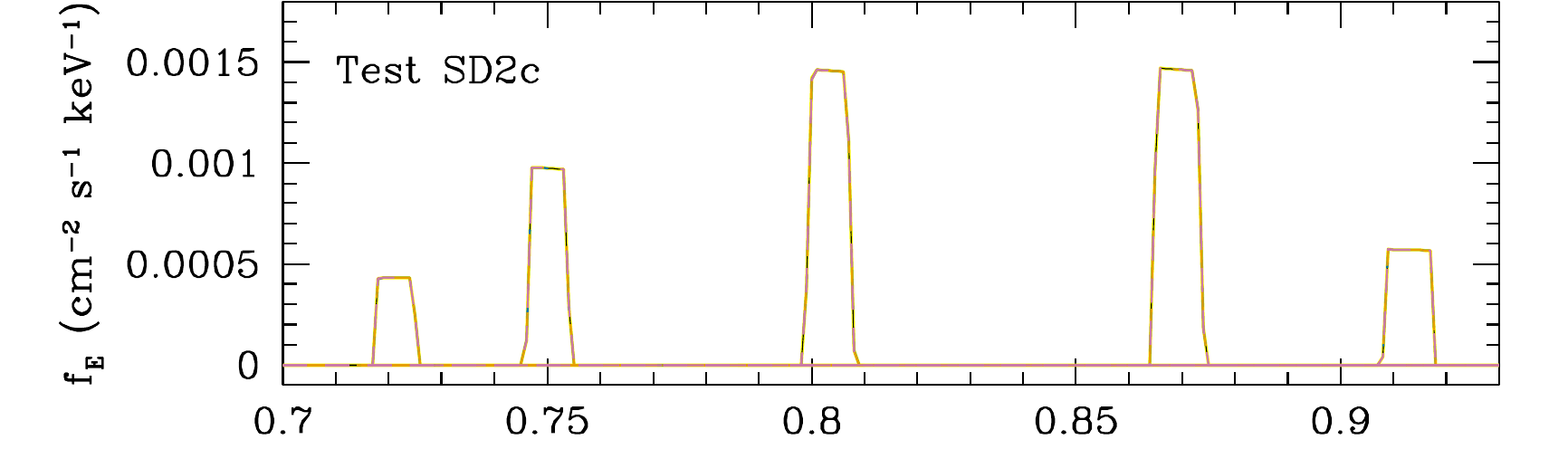}
  \includegraphics[clip, trim=4.2cm 19.2cm 1cm 3.7cm,angle=0,width=0.46\textwidth]{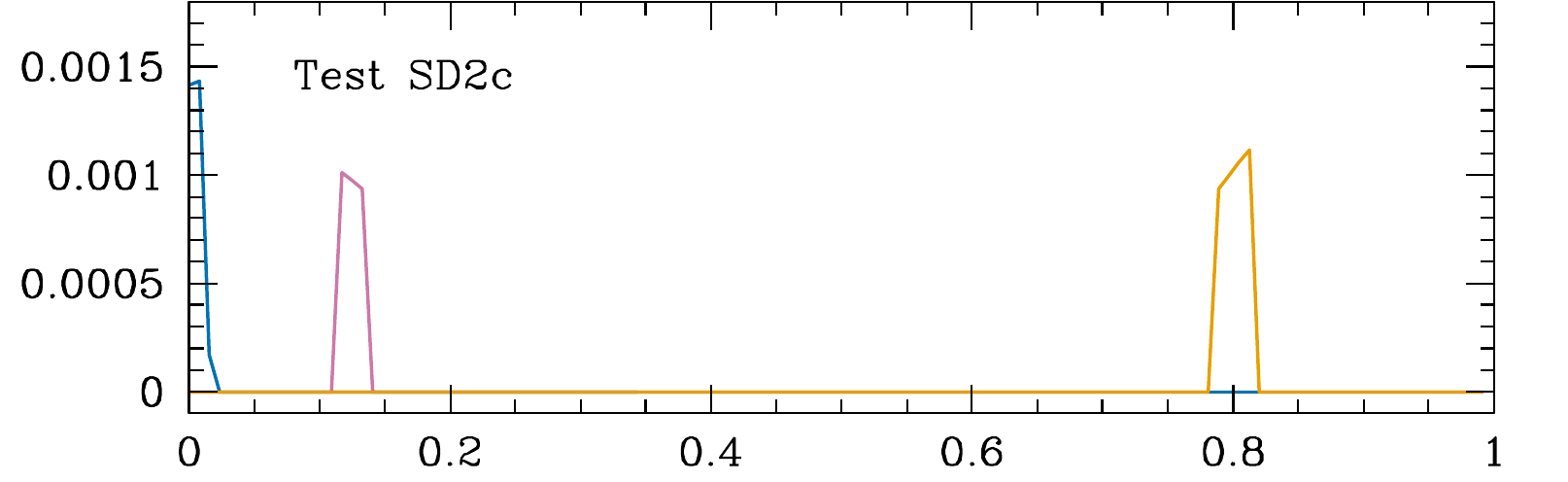}
  \includegraphics[clip, trim=3cm 18.5cm 1cm 3.7cm,angle=0,width=0.495\textwidth]{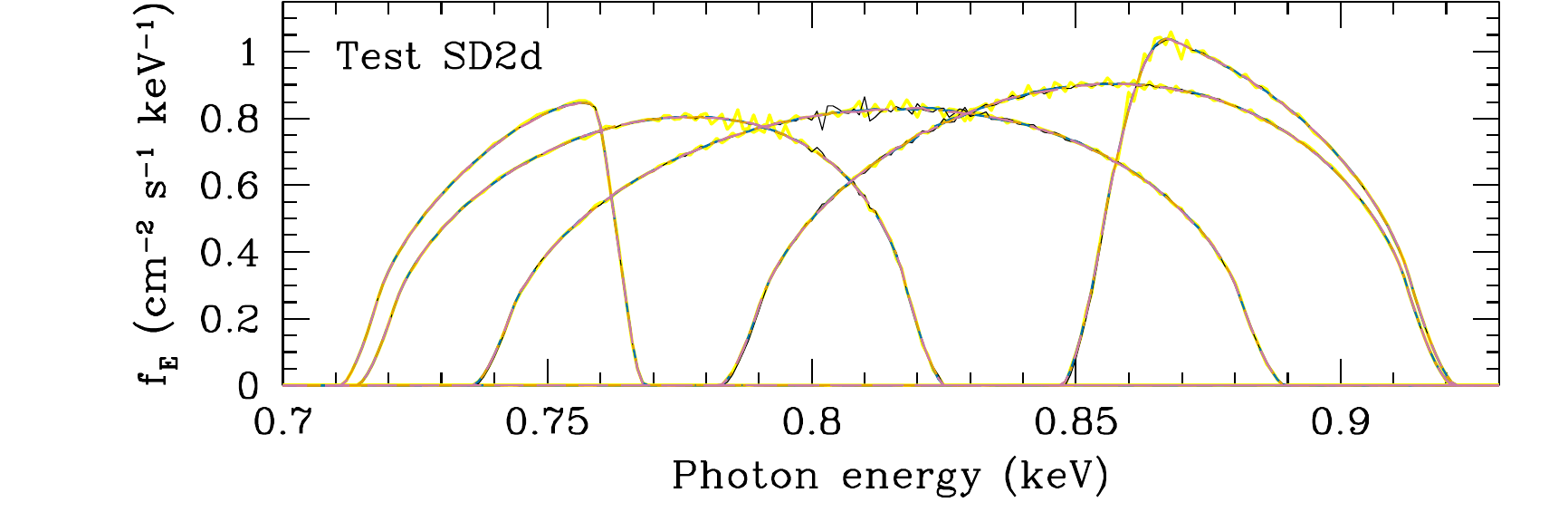}
  \includegraphics[clip, trim=4.2cm 18.5cm 1cm 3.7cm,angle=0,width=0.46\textwidth]{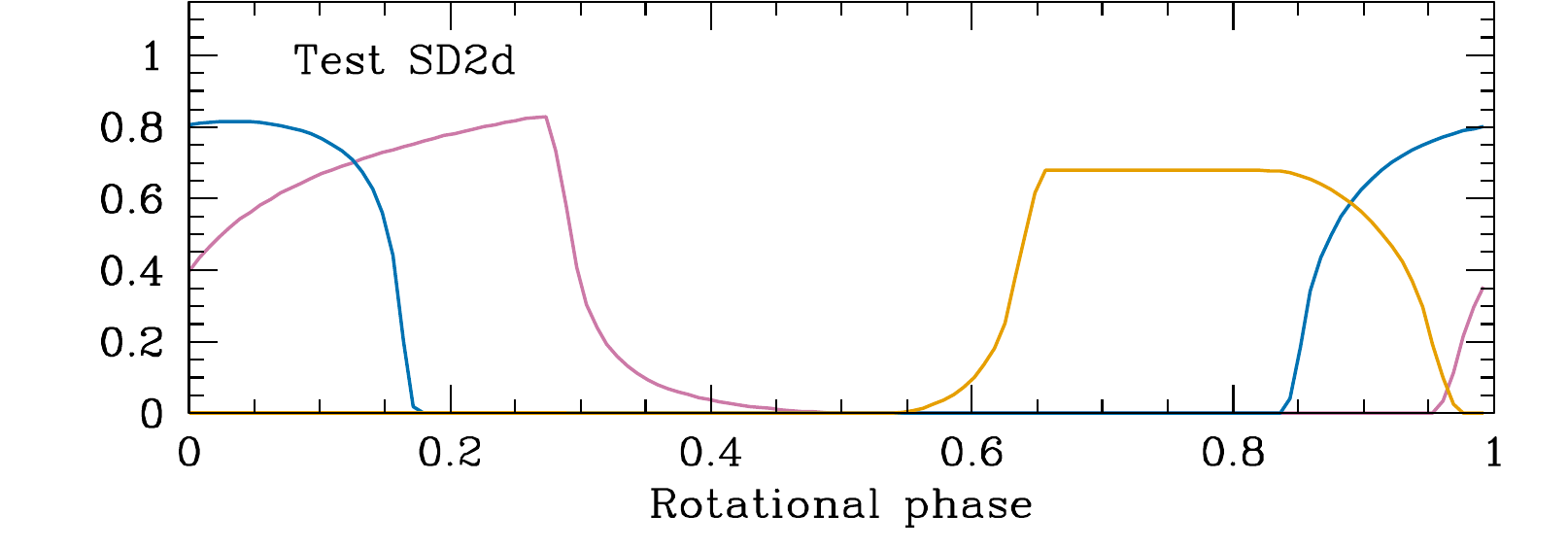}
  \caption{\textit{Left}: Results of the emission line tests SD2a--SD2d (from top to bottom, respectively) for five representative spin phases ($\phi=0.25$, $0.125$, $0$, $0.75$, and $0.875$, in order of increasing photon energy of the observed line) for the CU (black), IM (yellow), GSFC-M (orange), GSFC-S (blue), and AB (purple) codes.  The spikiness evident in the spectra at some phases for the CU and IM codes is produced due to inexact interpolation of the line profiles.  This interpolation problem is not an issue for realistic smooth spectra. \textit{Right}: For test SD2a and SD2b the monochromatic pulse profiles at energies $0.8094$\,keV (produced by the AB code and shown in purple at $\phi\approx 0.2$), $0.8096$\,keV (produced by the GSFC-S and shown in blue at $\phi\approx 0$), and $0.8098$\,keV (produced by the GSFC-M and shown in orange at $\phi\approx 0.8$) are shown. For tests SD2c and SD2d, the monochromatic profiles at energies 0.75\,keV (AB, purple), 0.8\,keV (GSFC-S, blue), and 0.9\,keV (GSFC-M, orange) are shown.
  Due to the narrow band nature of the emission lines, at a given photon energy in the observer rest frame the spot emission is only observed at some rotational phases.}
\label{fig:linetests}
\end{figure}

Figure~\ref{fig:linetests} shows a comparison of the emission line
outputs as a function of energy for five representative spin phases:
$\phi=0$, 0.125, 0.25, 0.75, and 0.875. It is apparent that the codes
are in good agreement, meaning that the emission lines are correctly
translated to the observer's frame. For Tests SD2a and SD2c, the
agreement between the line fluxes are exceptional, typically well
within $\sim$0.1\% aside from a small number of outliers ($\sim$10 out
of 25600 and 38400 values for SD2a and SD2c, respecitvely) for the CU
and IM codes. 

\section{Oblate Schwarzschild Code Verification Tests}
\label{sec:ostests}

Following the same strategy as in the S+D case, we have devised a series of comparison exercises to test different aspects of modeling hot spot emission in the OS approximation. These include consideration of both point-like (0.01 radian) and extended (1 radian) hot spots. In addition, for a subset of tests, we introduce non-isotropic surface emission using simple $\cos^2 \sigma$ and $\sin^2 \sigma$ beaming patterns. As in the S+D tests, we assume the same NS with $M=1.4$~\msol at a distance of $D=200$ pc, with a $kT=0.35$ keV Planck spectrum hot spot. Since the star is now oblate, the value for the NS radius we quote is the circumferential equatorial radius $R_{\rm eq}$. We again consider monochromatic pulse profiles at 1 keV in units of photons\,cm$^{-2}$\,s$^{-1}$\,keV$^{-1}$.
We note that for a subset of tests, we consider spin frequencies of 600\,Hz, which as pointed out in Section~\ref{sec:accuracy} falls in a regime where the OS approximation introduces errors at the level of a few percent. Nevertheless, for the purposes of the code verification comparisons, considering such rapid spin provides enhanced sensitivity to any discrepancies in the implementation of the stellar oblateness and special relativistic effects, because they are much more pronounced for faster spins.

\begin{figure}[t!]
\centering
  \includegraphics[clip, trim=5.5cm 5.5cm 1cm 2.5cm,angle=0,width=0.42\textwidth]{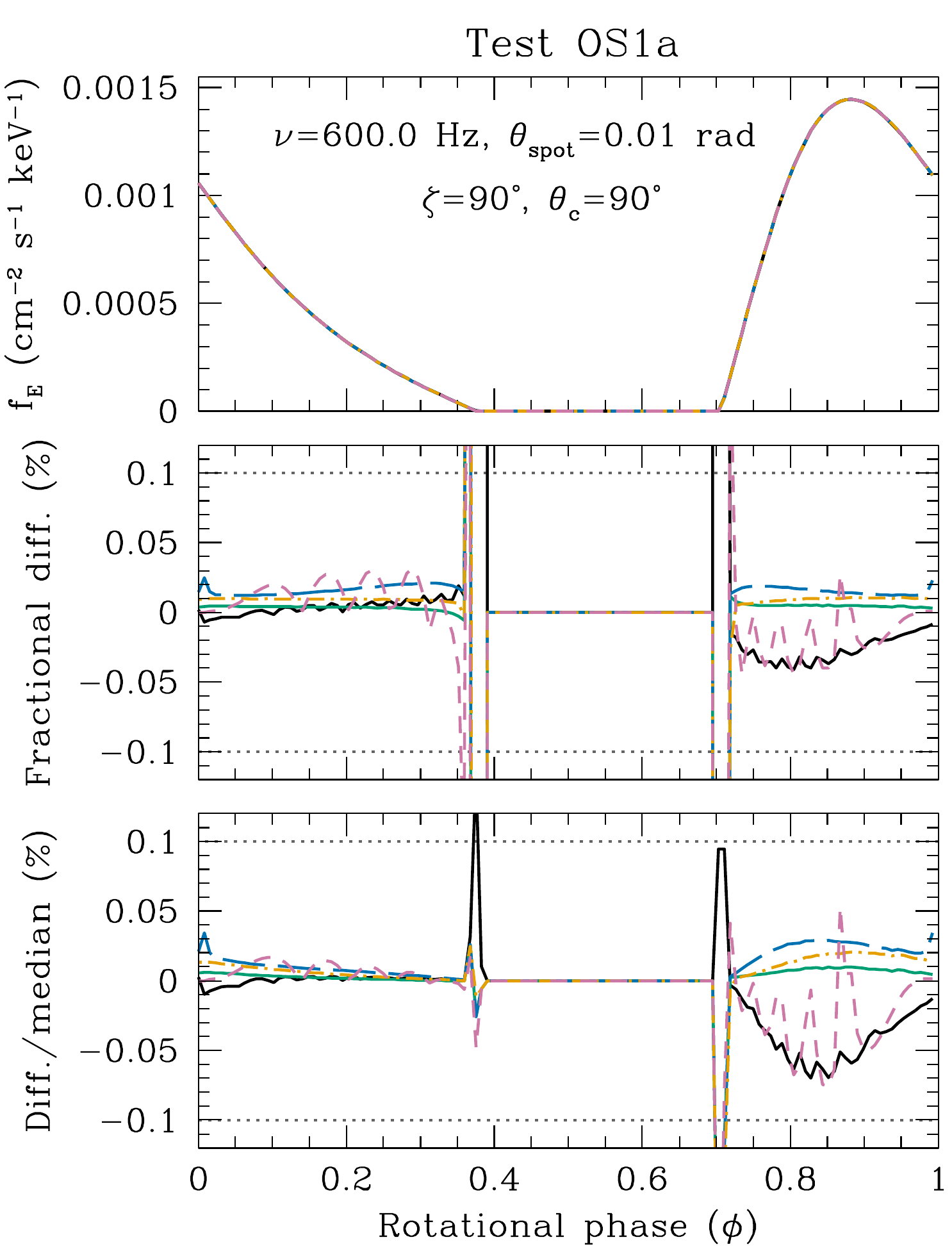}~~~
  \includegraphics[clip, trim=5.5cm 5.5cm 1cm 2.5cm,angle=0,width=0.42\textwidth]{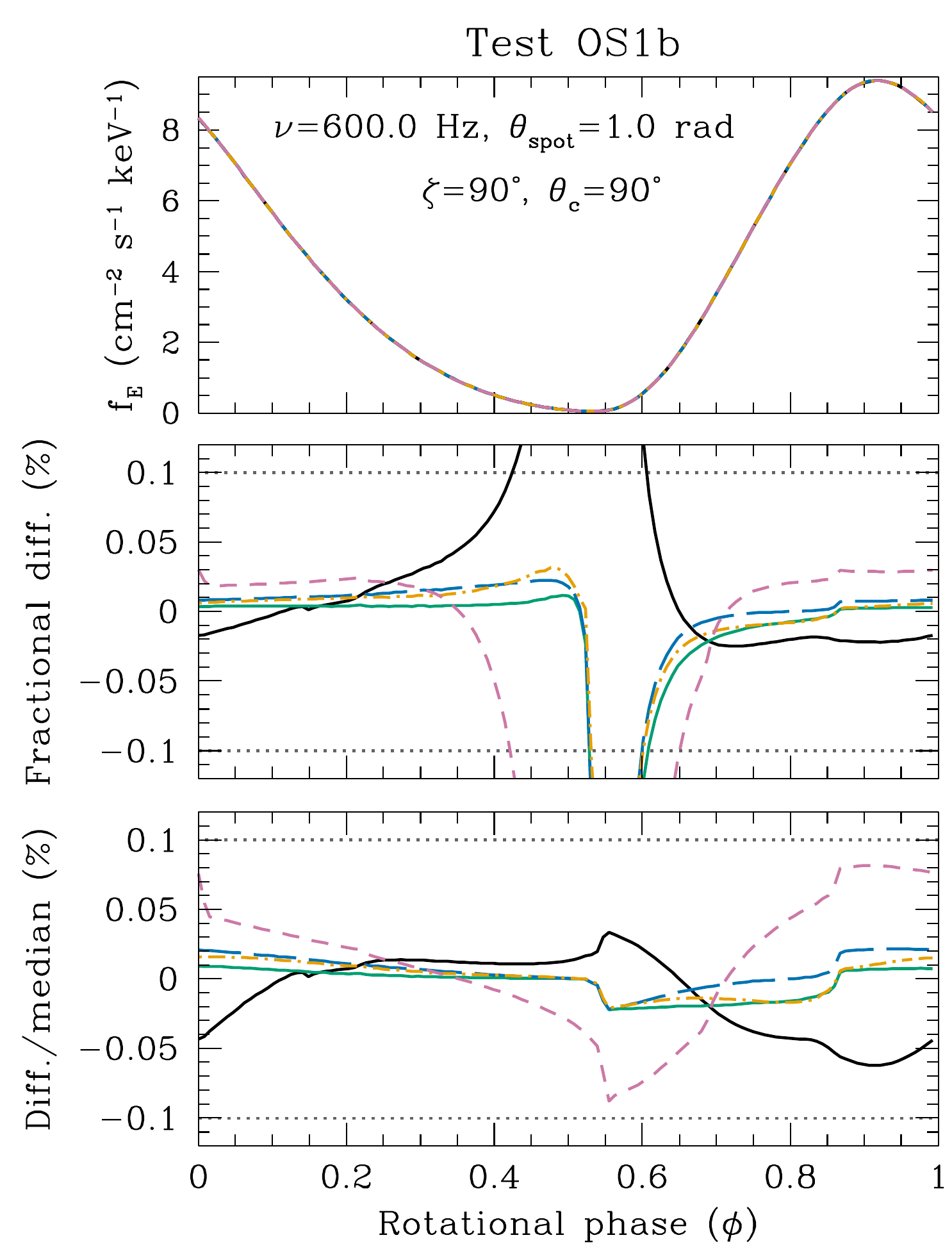}
  \caption{Comparisons of the synthetic pulse waveforms from Tests~OS1a and OS1b. The point-like ($\theta_{\rm spot}=0.01$ rad) and large spot ($\theta_{\rm spot}=1$ rad) are shown on the left and right, respectively. The top panel shows the pulse waveforms. The middle panel shows the fractional difference between the CU (black), GSFC-M (orange), GSFC-S (blue), Alberta (purple), and Amsterdam (green) fluxes and the IM flux at each phase bin, expressed as a percentage. In the bottom panel, the IM flux is subtracted from the other fluxes and the result is divided by the median IM flux over all phases. The pair of dotted lines show the target $\pm$0.1\% measurement precision.  Except near eclipses, where the flux is two orders of magnitude smaller than it is at the peak, the agreement between the codes is significantly better than the target precision.}
\label{fig:test_os1ab}
\end{figure}

\begin{figure}[t!]
\centering
  \includegraphics[clip, trim=5.5cm 5.5cm 1cm 2.5cm,angle=0,width=0.42\textwidth]{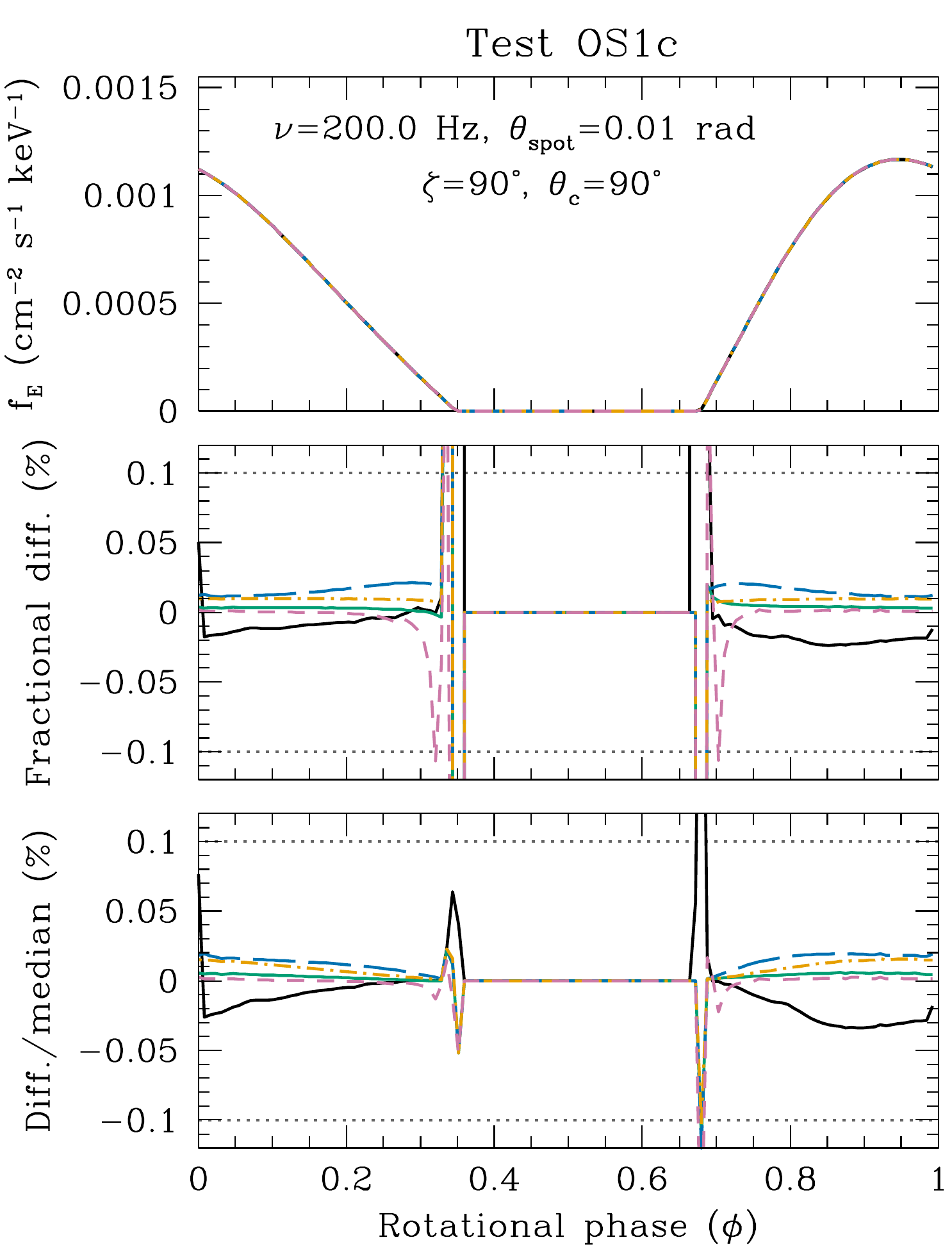}~~~
  \includegraphics[clip, trim=5.5cm 5.5cm 1cm 2.5cm,angle=0,width=0.42\textwidth]{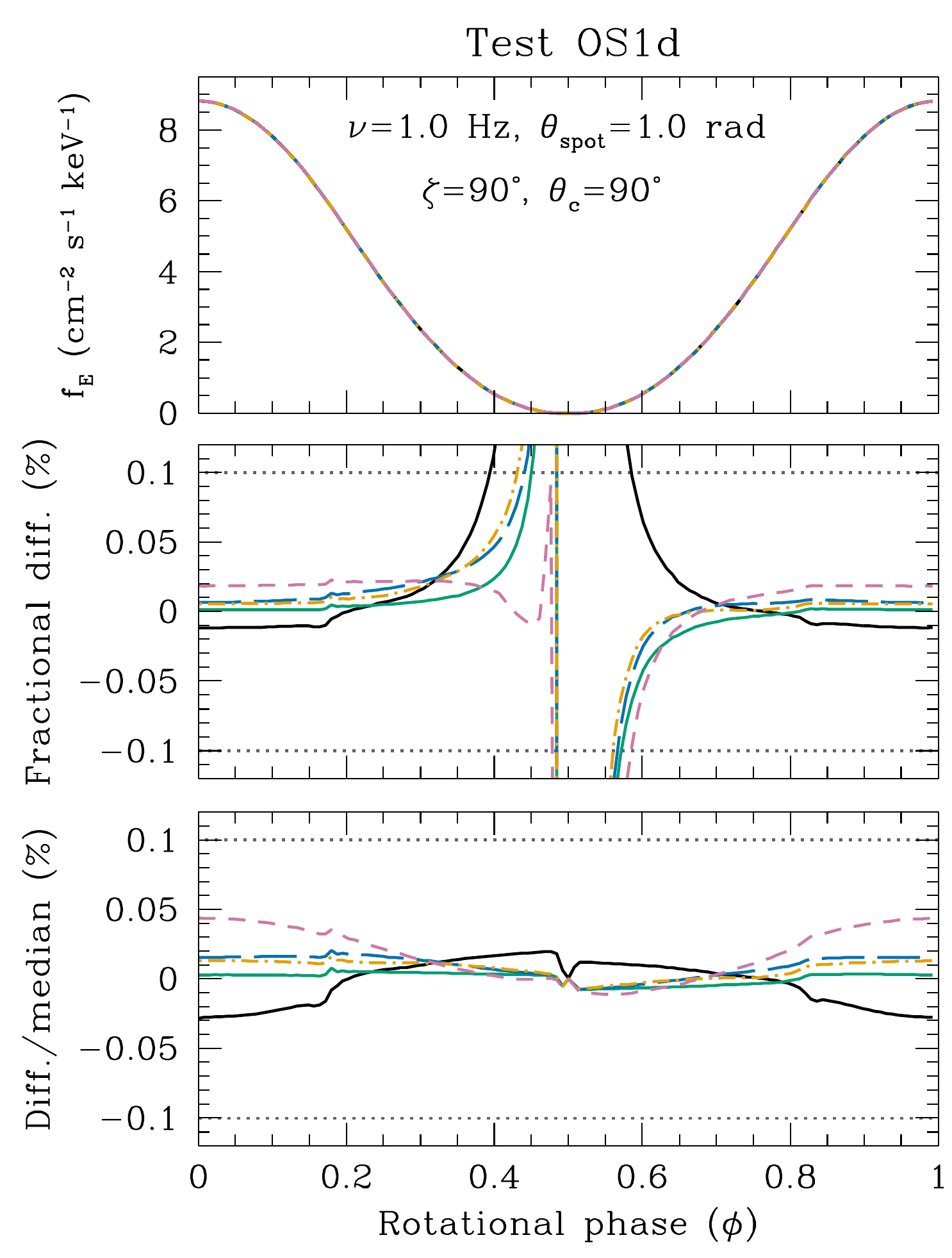}
  \caption{Same as Figure~\ref{fig:test_os1ab} but for tests OS1c and OS1d.}
\label{fig:test_os1cd}
\end{figure}

\begin{figure}[t!]
\centering
  \includegraphics[clip, trim=5.5cm 5.5cm 1cm 2.5cm,angle=0,width=0.42\textwidth]{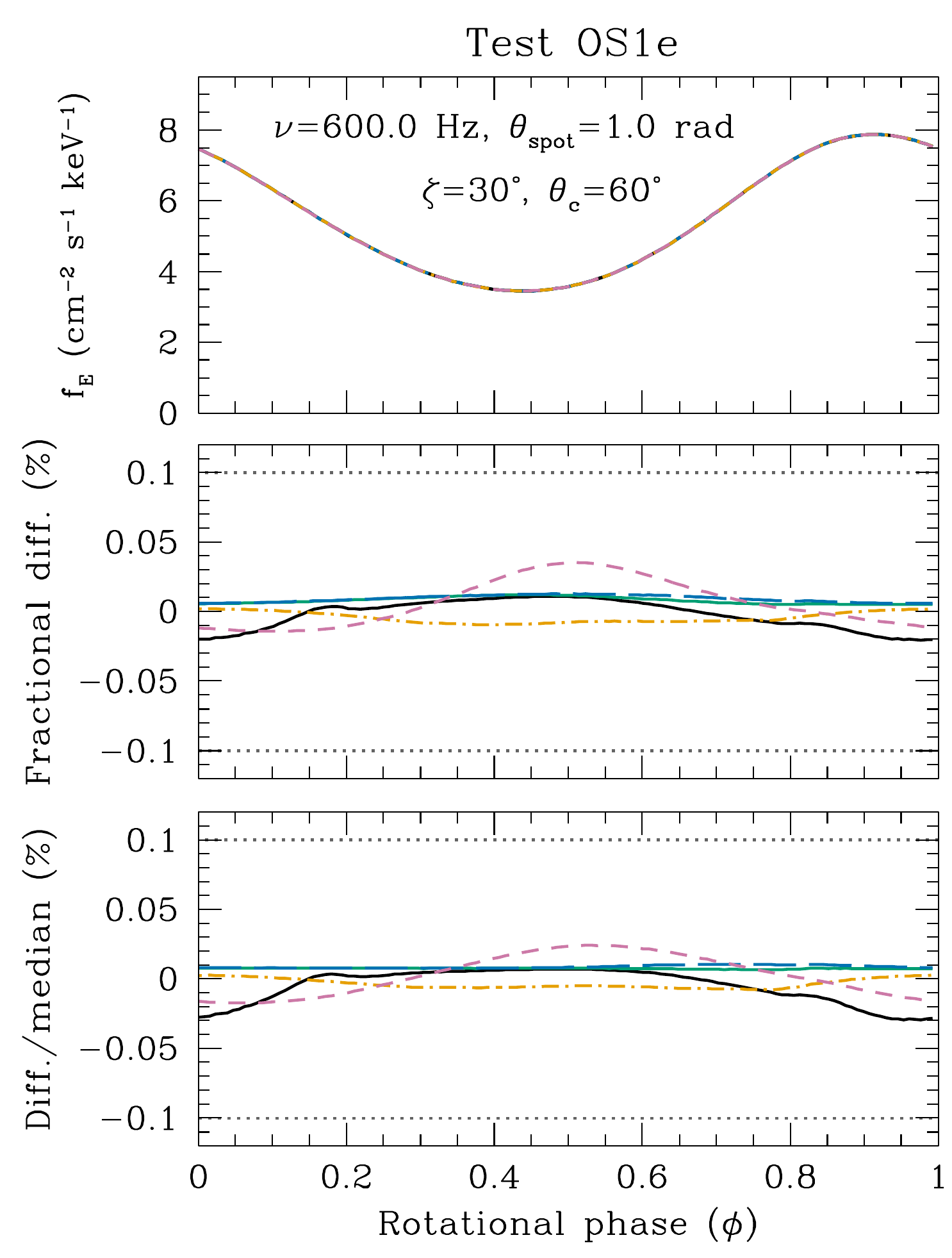}~~~
  \includegraphics[clip, trim=5.5cm 5.5cm 1cm 2.5cm,angle=0,width=0.42\textwidth]{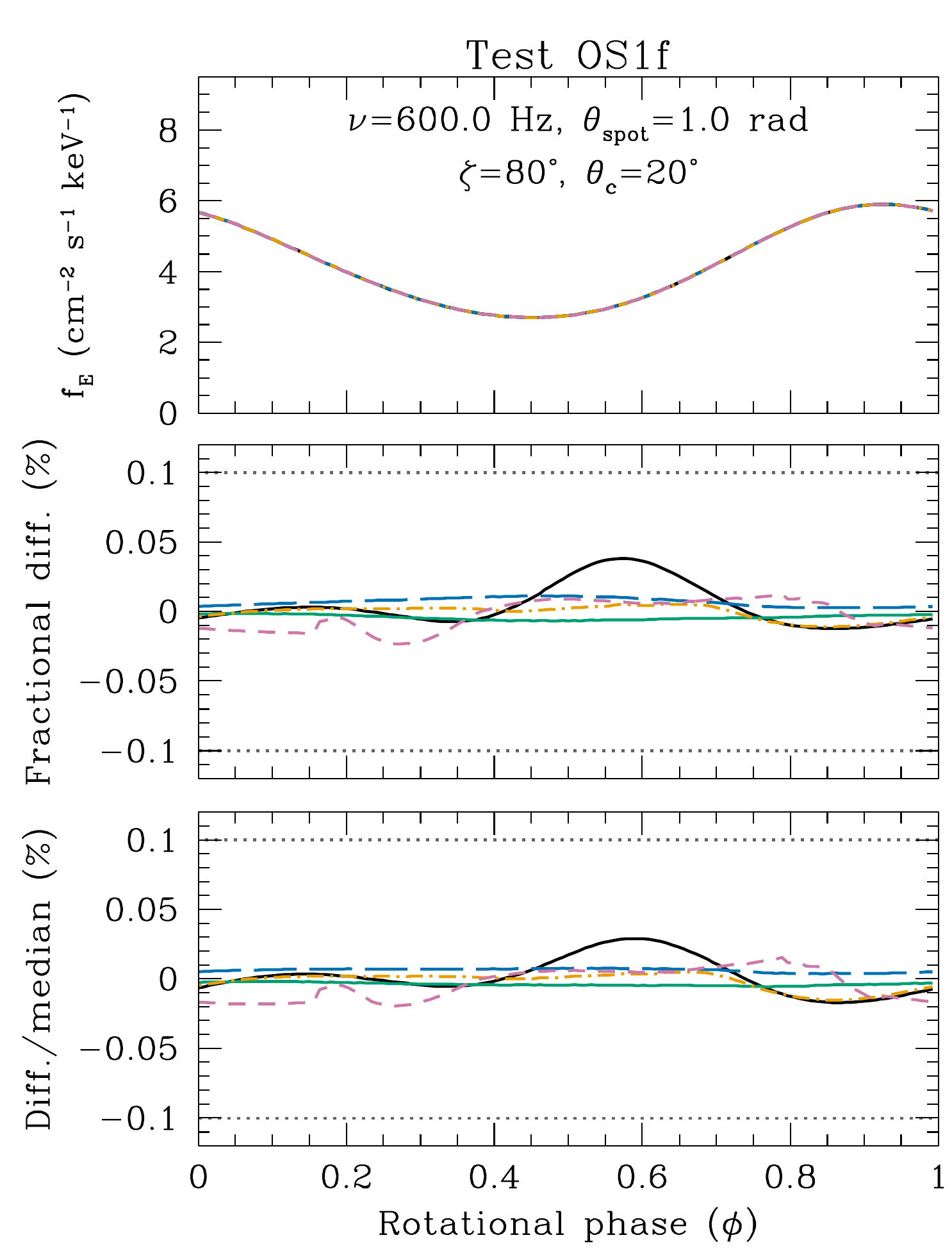}
  \caption{Same as Figure~\ref{fig:test_os1ab} but for tests OS1e and OS1f.}
\label{fig:test_os1ef}
\end{figure}

\begin{figure}[t!]
\begin{center}
  \includegraphics[clip, trim=5.5cm 5.5cm 1cm 2.5cm,angle=0,width=0.42\textwidth]{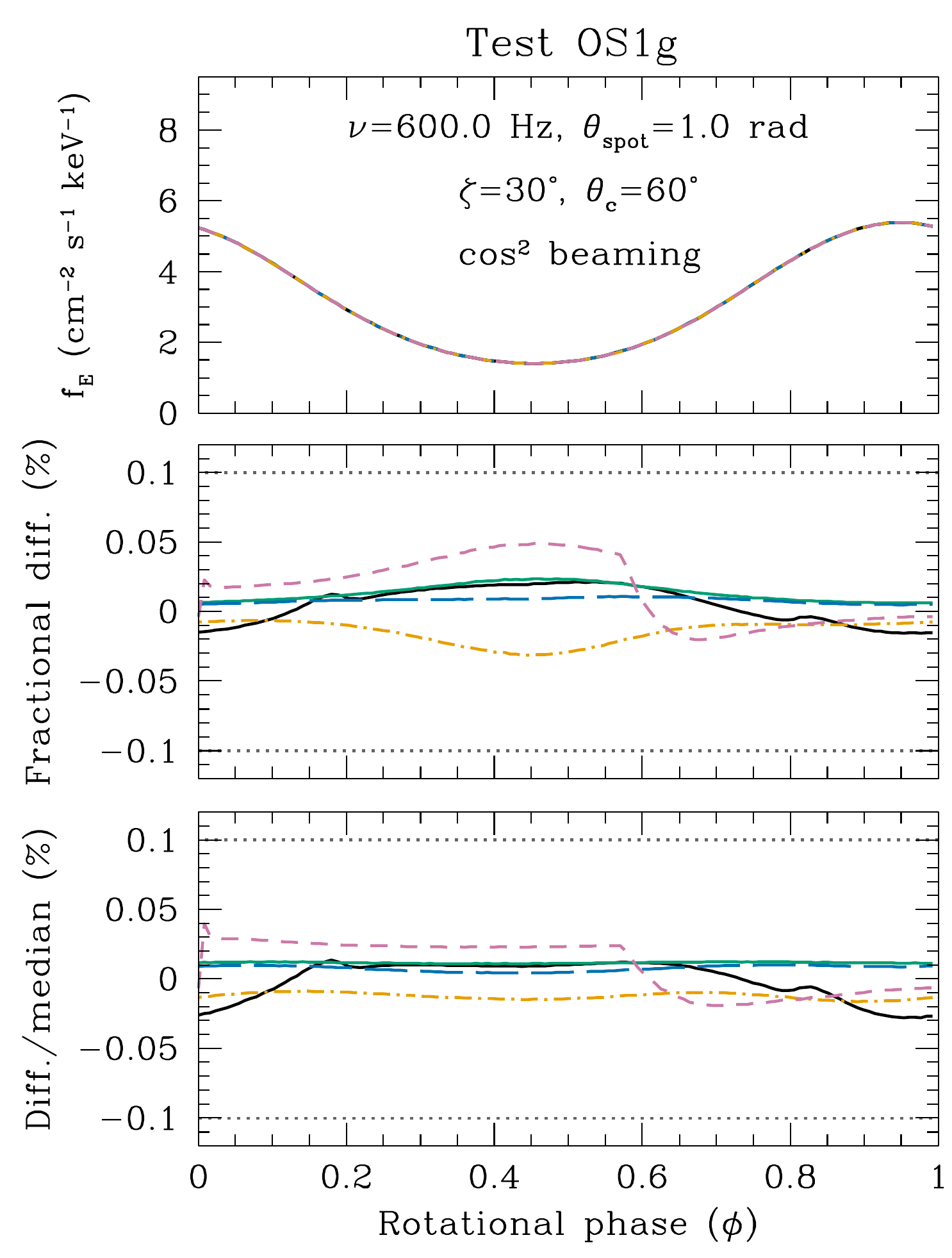}~~~
  \includegraphics[clip, trim=5.5cm 5.5cm 1cm 2.5cm,angle=0,width=0.42\textwidth]{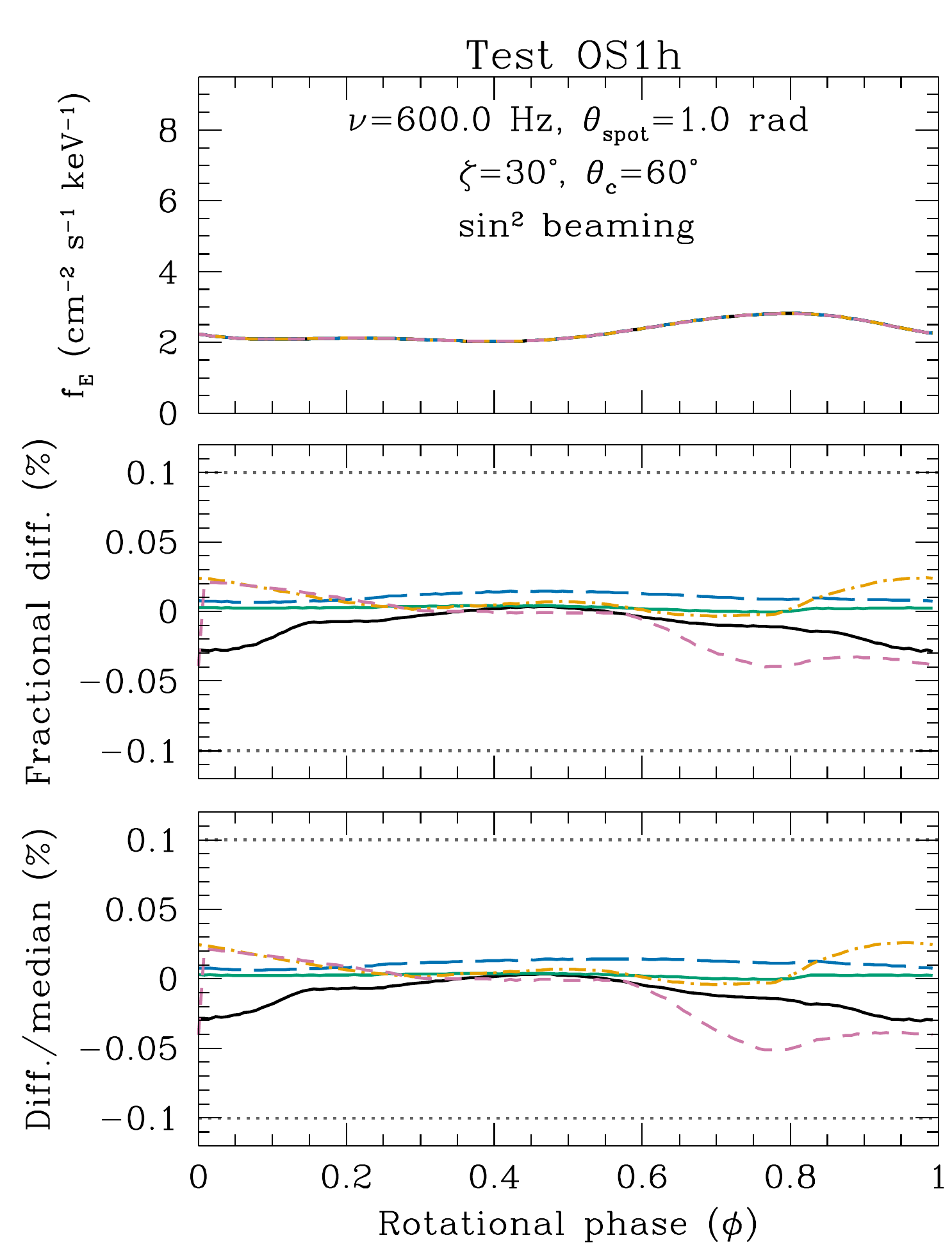}
  \caption{Same as Figure~\ref{fig:test_os1ab} but for tests OS1g and OShf.}
\label{fig:test_os1gh}
\end{center}
\end{figure}

\begin{figure}[t!]
\begin{center}
  \includegraphics[clip, trim=5.5cm 5.5cm 1cm 2.5cm,angle=0,width=0.42\textwidth]{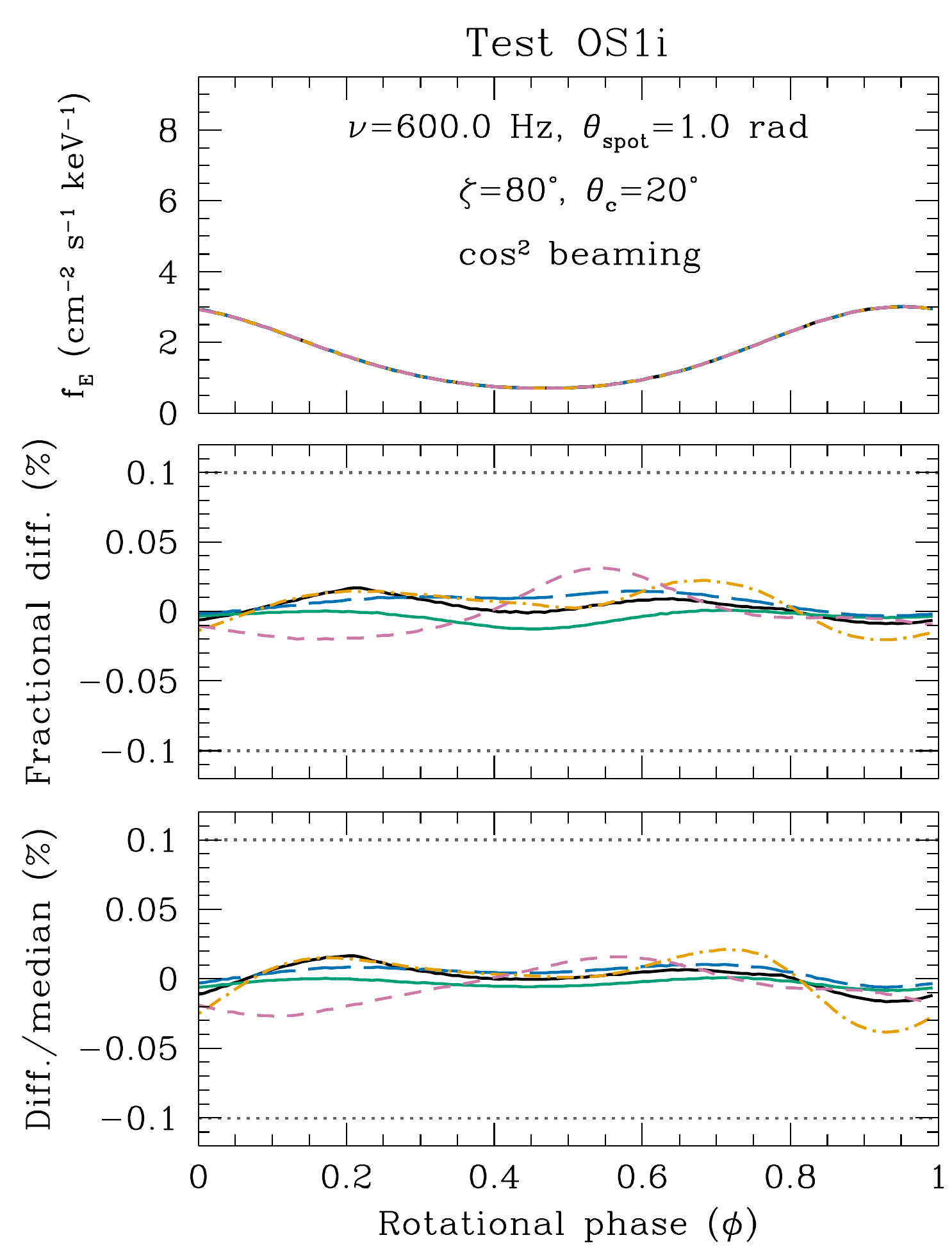}~~~
  \includegraphics[clip, trim=5.5cm 5.5cm 1cm 2.5cm,angle=0,width=0.42\textwidth]{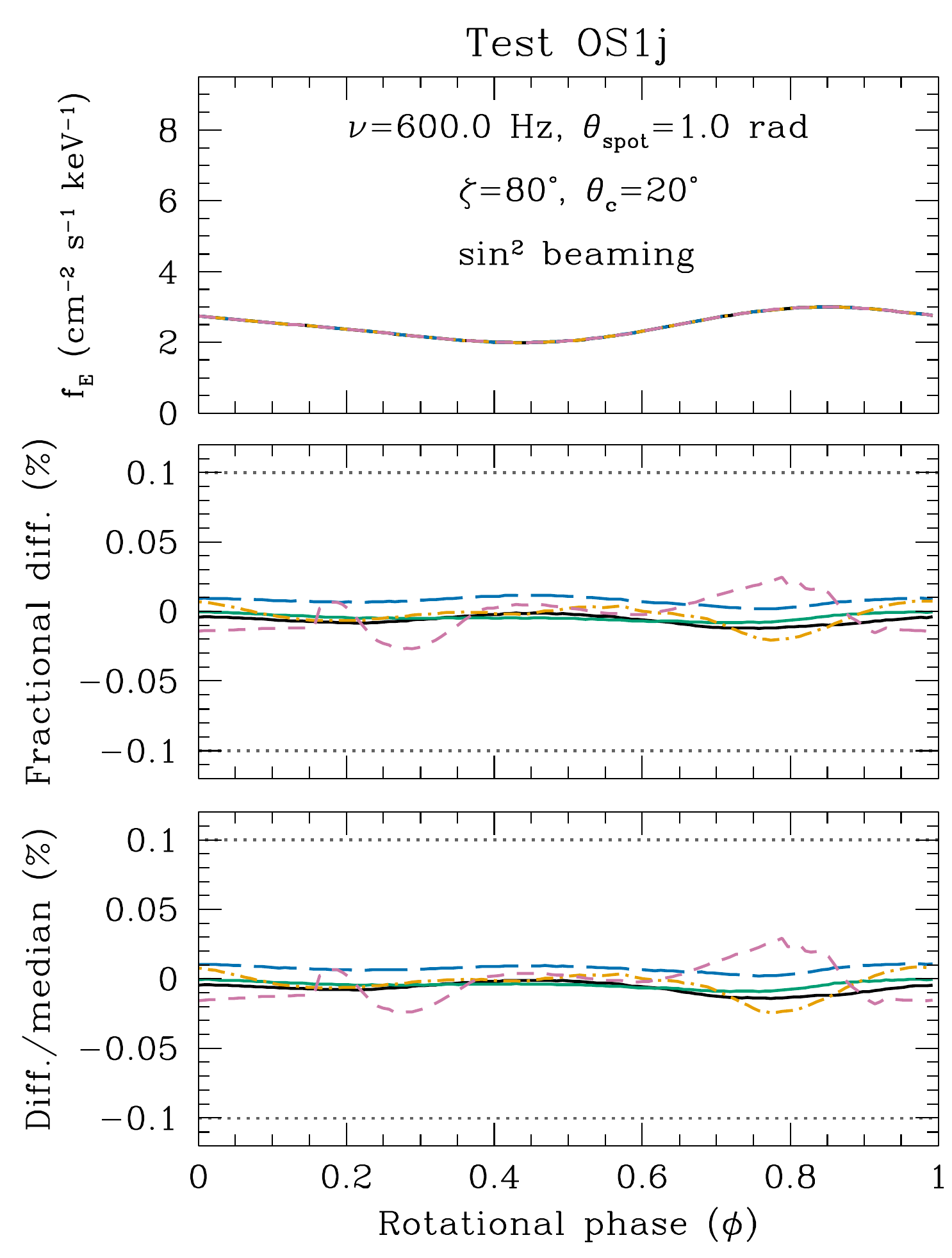}
  \caption{Same as Figure~\ref{fig:test_os1ab} but for tests OS1i and OS1j.}
\label{fig:test_os1ij}
\end{center}
\end{figure}

\begin{figure}[t!]
\centering
  \includegraphics[clip, trim=3cm 19.2cm 1cm 3.7cm,angle=0,width=0.495\textwidth]{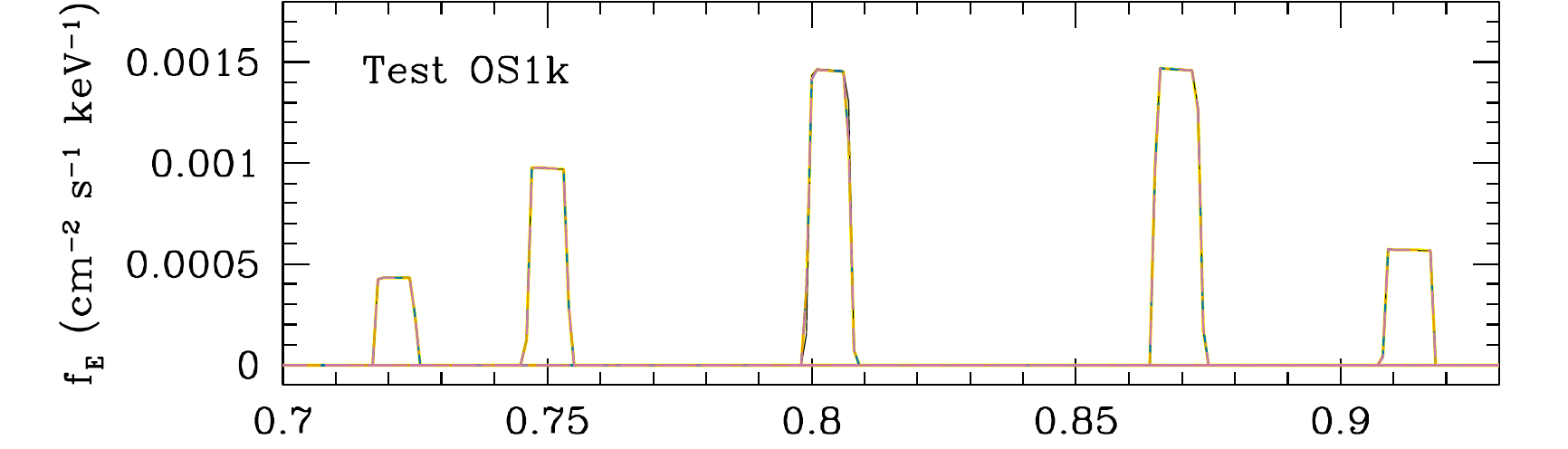}
  \includegraphics[clip, trim=4.2cm 19.2cm 1cm 3.7cm,angle=0,width=0.46\textwidth]{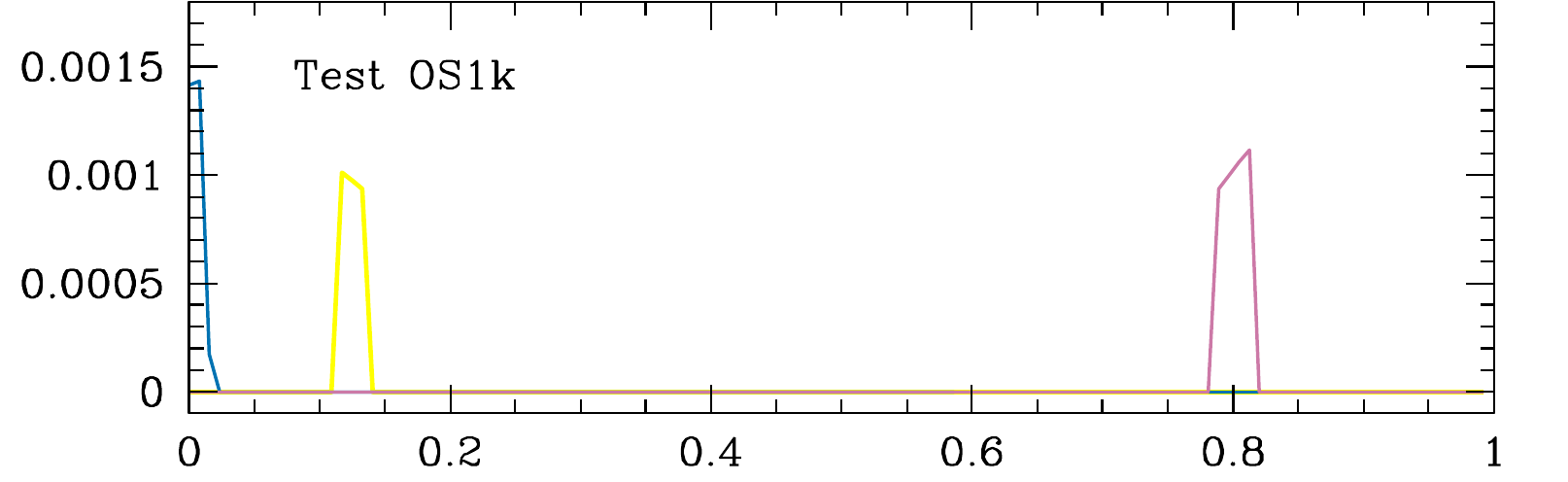}
  \includegraphics[clip, trim=3cm 18.5cm 1cm 3.7cm,angle=0,width=0.495\textwidth]{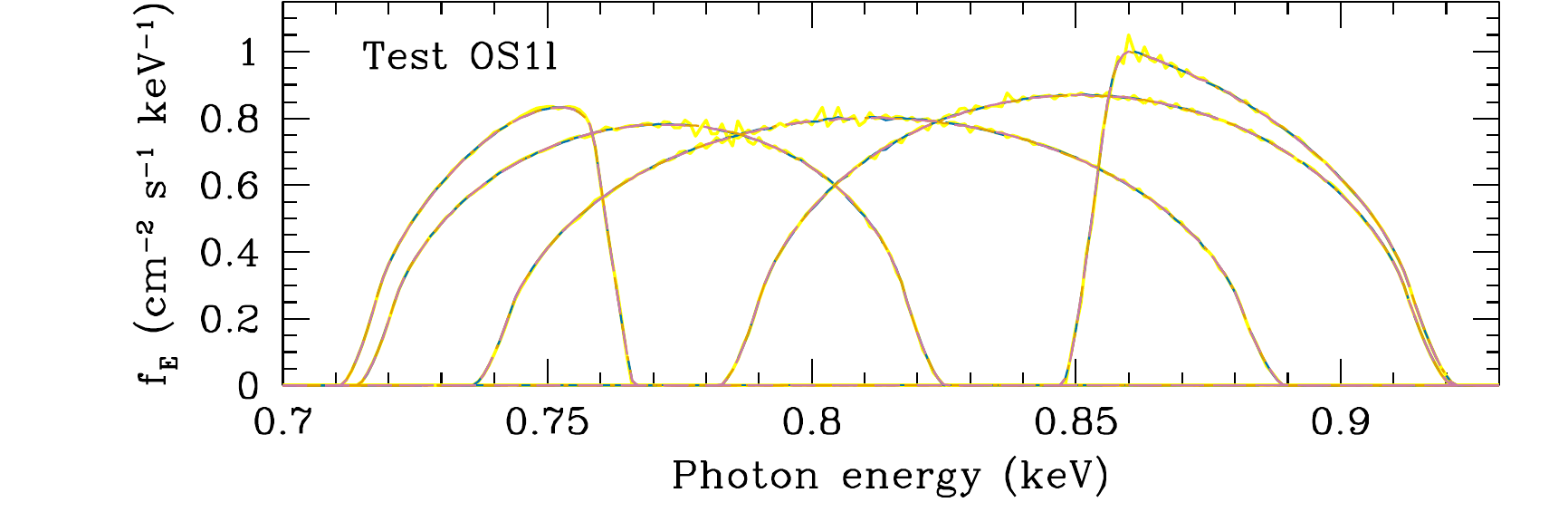}
  \includegraphics[clip, trim=4.2cm 18.5cm 1cm 3.7cm,angle=0,width=0.46\textwidth]{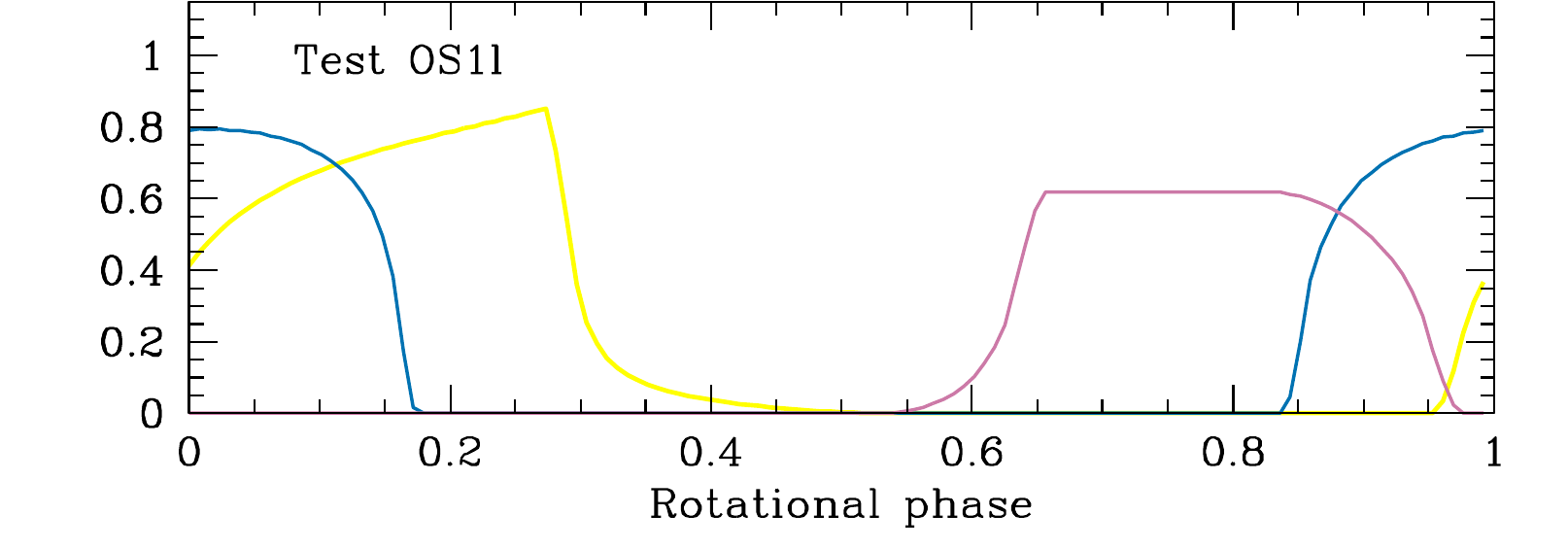}
  \caption{\textit{Left}: Results of the emission line tests OS1k and OS1l (from top to bottom, respectively) for five representative spin phases ($\phi=0.25$, 0.125, 0, 0.75, and 0.875, in order of increasing photon energy of the observed line) for the CU (black), IM (yellow), GSFC-M (orange), GSFC-S (blue), and AB (purple) codes.  The spikiness evident in the spectra at some phases for the CU and IM codes is produced due to inexact interpolation of the line profiles.  \textit{Right}: Monochromatic pulse profiles for the OS1k and OS1l tests at energies 0.75\,keV at $\phi\approx0.125$ (from IM code, marked in yellow), 0.8\,keV at $\phi\approx0.0$ (from GSFC-S code, marked in blue), and 0.9\,keV at $\phi\approx0.8$ (from AB code, marked in purple). Due to the narrow band nature of the emission lines, at a given photon energy in the observer rest frame the spot emission is only observed at some rotational phases. }
\label{fig:oslinetests}
\end{figure}

The setup of the series of tests is summarized in Tables~\ref{table:ostest1af} and \ref{table:ostest1gl}, while the results of the OS code comparisons are illustrated in Figures~\ref{fig:test_os1ab}--\ref{fig:oslinetests}. The IM code was again used as a reference against which the other codes were compared. In these OS tests, results from the AMS codes (see Appendix~\ref{sec:codes}) are also included; in addition to the ``star-to-observer'' ray-tracing technique described here, the AMS codes can also employ ``observer-to-star'' (i.e., image plane) ray-tracing based on the prescription from \citet{2014ApJ...792...87P}. For the intended purposes the two approaches to ray-tracing produce results that are virtually indistinguishable. As with the S+D tests, for most rotational phases, the deviations in computed photon flux are comfortably below the 0.1\% requirement. The exceptions are phase bins around the ingress and egress as the hot spot is occulted by the neutron star. However, in these cases the photon fluxes are minuscule, being more than two orders of magnitude smaller than the flux around the pulse maximum.  Based on this, we deem these discrepancies to be unimportant for practical purposes.  Specifically, in actual observational data, the measured flux at pulse minimum will typically be dominated by other source emission components or non-source background. Even in the event of negligible background, the measurement uncertainty of the low count rate at the spot ingress/egress would dominate over a $\sim$0.1\% numerical  imprecision. Finally, the likelihood functions used for parameter estimation based on the pulse profile modeling technique are generally insensitive to error at phases in the vicinity of ingress/egress.


\begin{table}[h]
\begin{center}
\footnotesize{
  \begin{tabular}{lcccccc}
    \hline
    \emph{Quantity} & \emph{Test OS1a} & \emph{Test OS1b} & \emph{Test OS1c} & \emph{Test OS1d} & \emph{Test OS1e} & \emph{Test OS1f}\\
    \hline
    Number of hot spots              & 1   & 1   & 1    & 1 & 1 & 1\\
    Colatitude of spot center (\degr) & 90  & 90  & 90   & 90 & 60 & 20\\
    Angular radius of hot spot (rad)  & 0.01 & 1.0 & 0.01 & 1.0 & 1.0 & 1.0\\
    Colatitude of observer (\degr) & 90  & 90  & 90   & 90 & 30 & 80\\
    Neutron star mass (\msol)     & 1.4 & 1.4 & 1.4  & 1.4 & 1.4 & 1.4\\
    Neutron star equatorial radius (km)         & 12 & 12   & 12   & 12 & 12 & 12\\
    $\nu$ at infinity (Hz) &  600 &  600   & 200  & 1 & 600 & 600\\
    Spectrum of emission             & Planck & Planck & Planck & Planck & Planck & Planck \\
    Beaming of emission              & iso & iso & iso & iso & iso & iso\\
    Temperature of emission (keV)    & 0.35 & 0.35   & 0.35 & 0.35 & 0.35 & 0.35\\
    \hline
  \end{tabular}
  \caption{Parameter values for Waveform Test OS1 a-f.}
    \label{table:ostest1af}}
\end{center}  
\end{table}


\begin{table}[h]
\begin{center}
\footnotesize{
  \begin{tabular}{lcccccc}
    \hline
    \emph{Quantity} & \emph{Test OS1g} & \emph{Test OS1h} & \emph{Test OS1i} & \emph{Test OS1j} & \emph{Test OS1k} & \emph{Test OS1l}\\
    \hline
    Number of hot spots              & 1   & 1   & 1    & 1 & 1 & 1\\
    Colatitude of spot center (\degr) & 60  & 60  & 20   & 20 & 90 & 90\\
    Angular radius of hot spot (rad)  & 1.0 & 1.0 & 1.0 & 1.0 & 1.0 & 0.01\\
    Colatitude of observer (\degr) & 30  & 30  & 80   & 80 & 90 & 90\\
    Neutron star mass (\msol)     & 1.4 & 1.4 & 1.4  & 1.4 & 1.4 & 1.4\\
    Neutron star equatorial radius (km)         & 12 & 12   & 12   & 12 & 12 & 12\\
    $\nu$ at infinity (Hz) &  600 &  600   & 600  & 600 & 600 & 600\\
    Spectrum of emission             & Planck & Planck & Planck & Planck & line Planck & line Planck \\
    Beaming of emission              & $\cos^2\sigma$ & $\sin^2\sigma$ & $\cos^2\sigma$ & $\sin^2\sigma$ & iso & iso\\
    Temperature of emission (keV)    & 0.35 & 0.35   & 0.35 & 0.35 & 0.35 & 0.35\\
    \hline
  \end{tabular}
  \caption{Parameter values for Waveform Test OS1 g-l.
}
    \label{table:ostest1gl}}
\end{center}  
\end{table}

\section{Conclusions}
\label{sec:conclusions}
We have described the model of hot spot emission from a rapidly rotating neutron star that we intend to apply to \textsl{NICER} X-ray data of MSPs.
We have presented the results of the first direct comparison between independently developed codes to allow a crucial consistency check of the calculations that go into a model of emission from a rapidly rotating neutron star in both the S+D and OS approximations. We find that the outputs of the codes are in excellent agreement with one another and with several analytical and semi-analytical results, with fractional differences of $\lesssim$0.1\%. In addition, we obtain consistent results with the ``star-to-observer'' and ``observer-to-star'' (i.e., image plane) ray-tracing techniques (see Appendix~\ref{sec:codes}). The  set  of  verified,  high-precision  reference  synthetic pulse  profiles for both the S+D and OS comparisons is provided  to  the  community as supplementary material to this article to  facilitate testing  of  other  independently  developed codes.

A crucial aspect of the comparisons presented in this work is that the synthetic pulse profiles were generated with what we call ``practical'' versions of the codes, meaning that they strike an optimal balance between providing accuracy at a level better than 0.1\% and having short execution times.  While, in principle, we could obtain substantially better agreement by increasing the number of resolution elements in the codes, this would come at the expense of extra computational time. For the practical applications of these codes in parameter estimation analyses, for each sampled combination of parameters it is necessary to generate a synthetic pulse profile for hundreds of detector energy channels. With this in mind, the numerical performance of the codes described here has been optimized so that the computation of a synthetic pulse profile at a given energy is executed in much less than 1\,s on a single processor core (see Appendix~\ref{sec:codes} for details). As demonstrated in \cite{miller19} and \cite{riley19}, the consistency between the codes and their computational efficacy makes them well-suited for use in the large-scale statistical sampling runs required to obtain estimates on the $M$-$R$ relation and the dense matter EoS.

\appendix


\section{Parameter Notation} 
\label{sec:parameters}

Here we provide a summary table that defines the notation for the numerous symbols used throughout this paper, which for many parameters differs from previous publications. For convenience, Table \ref{table:parameters} lists each symbol and a brief definition.

\begin{table}[h]
\begin{center}
  \caption{Notation of the various parameters used in this paper.
  }
   \label{table:parameters}
  \begin{tabular}{ll}
    \hline
    Symbol & Description \\
    \hline
$\vec{n}$ \dotfill & Vector normal to the stellar surface\\
    $\alpha$ \dotfill &  Angle between the radial direction and the initial direction of the light ray\\
	$\psi$ \dotfill & Deflection angle between the initial and final photon directions\\
    $\Delta t$ \dotfill & Difference in elapsed coordinate time between outgoing ray and reference ray\\
	$\zeta$ \dotfill & Observer colatitude, measured from the spin axis\\
    $\phi$ \dotfill & Stellar rotational phase \\
    $\phi_{\rm obs}$ \dotfill & Observed rotational phase \\
    $\theta$ \dotfill & Colatitude of a point/small patch on the star, measured from the spin axis\\
    $\theta_c$ \dotfill & Colatitude of hot spot center\\
    $\gamma$ \dotfill & Lorentz factor \\
    $\delta$ \dotfill & Relativistic Doppler factor \\
    $v$      \dotfill & Speed of the surface as measured by local static observer  \\
    $\beta \equiv v/c$ \dotfill & Dimensionless speed \\
    $\xi$ \dotfill & Angle between emission direction and direction of motion of stellar surface\\
    $R$ \dotfill & Neutron star radius at latitude of emitting area \\
    $M$ \dotfill & Neutron star mass\\
    $R_S\equiv 2GM/c^2$ \dotfill & Schwarzschild radius\\
     $\nu$ \dotfill & Spin frequency of the star, as measured by an observer at infinity. Units of Hz\\
    $\Omega \equiv 2 \pi \nu$ \dotfill & Angular spin frequency \\
    $\Delta\theta$, $\theta_{\rm spot}$ \dotfill & Angular radius of spot\\
    $b$ \dotfill & Light ray impact parameter\\
    $g_0$ \dotfill & Surface gravity of spherical neutron star \\
    $E$ \dotfill & Photon energy \\
    $I_E$ \dotfill & Specific intensity of radiation at photon energy $E$ \\
    $F$ \dotfill& Photon flux incident at distant observer \\
    $D$ \dotfill & Distance between the star and observer\\
    \hline
    OS Approximation & \\
    \hline
    $\sigma$ \dotfill & Angle between the local surface normal and the initial photon direction \\
    $\tau$ \dotfill & Angle between radial direction and local surface normal \\
    $\lambda$ \dotfill & Local azimuthal angle \\
    $R_{\rm eq}$ \dotfill & Equatorial circumferential radius of an oblate neutron star \\
    $g(\theta)$ \dotfill & Surface gravity as a function of colatitude for an oblate neutron star \\
    \hline
  \end{tabular}
\end{center}  
\end{table}

\section{Code Descriptions}
\label{sec:codes}

We now briefly describe each of the codes that were used for the pulse-waveform comparisons described in Section~\ref{sec:verification}. 

\subsection{The Columbia Code}\label{app: CU code}
The Schwarzschild+Doppler code used by the Columbia University group (hereafter the CU code) is described in \citet{2007ApJ...670..668B,2008ApJ...689..407B}. The code is one of the first to consider a hydrogen atmosphere model to describe the spectrum and beaming pattern of neutron star emission in the context of pulse profile fitting.  A variant of the code implements the light bending approximations by \citet{2002ApJ...566L..85B}, which can be useful in analyses for which several-percent precision is adequate. However, for the tests described below, the correct light deflection formula is used.  The code has been developed in C++ and makes use of integration and interpolation routines from the third edition of ``Numerical Recipes'' \citep{2002nrca.book.....P}. The CU code has been successfully applied to \textit{XMM-Newton} data of the nearest rotation-powered millisecond pulsars, PSR~J0437$-$4715 \citep{2013ApJ...762...96B}, PSR~J0030+0451
\citep{2009ApJ...703.1557B}, and PSR~J2124--3358 \citep{2008ApJ...689..407B} to produce crude constraints on the NS $M-R$ relation.
    
The CU code is optimized to be both accurate and fast. In order to decrease computational cost, the light bending integral, the lensing factor, and the time delay integral are pre-computed on a fine grid in $\alpha$ based on the input $M$ and $R$. The resulting look-up tables are used to obtain the necessary values via quadratic interpolation when generating the output pulse waveforms. For a 360 point grid in $\alpha$ this approach still maintains high accuracy ($\lesssim 0.1\%$ errors in the deflection angle, time delay, and lensing factor). The light bending and travel time delay integral are computed with the variable substitution $x=\sqrt{1-R/r}$, which avoids the integrand divergences in Equation~(\ref{eq:defl}) and Equation~(\ref{eq:delay}) as $r\to R^{+}$ and $\alpha\to\pi/2$, and makes the integration domain compact.

The hot spot on the stellar surface is represented by a grid of small surface elements in colatitude $\theta_c$ and longitude $\phi$ that can be resized as needed. For a circular hot spot, the code determines whether a surface element falls within the boundaries of the hot spot. For each surface element in the spot, the CU code determines the value of $\psi$ via Equation~(\ref{eq:defl}) and obtains the corresponding $\alpha$, lensing factor ${\rm d}\cos\alpha/{\rm d}\cos\psi$, and travel time delay via interpolation from the pre-computed lookup tables. This procedure is repeated for the entire range of spin phases and the observed flux is computed using Equation~(\ref{eq:flux}). The final observed pulse waveform is generated after correcting for the phase shift caused by travel time differences.

An additional boost in speed is gained by taking advantage of the fact that at each slice in co-latitude, all surface elements with the same temperature produce the same pulse waveform but are shifted in phase. Thus, at a given co-latitude, it is only necessary to compute a single pulse waveform for one surface element and then shifting in phase and summing all pulse waveforms.

\subsection{The GSFC codes}

The waveform codes developed at GSFC (Strohmayer and Mahmoodifar, hereafter GSFC-S and -M) are based on initial implementations of the Schwarzschild+Doppler approximation, subsequently generalized to include the Oblate Schwarzchild approximation. Here we highlight several aspects of the GSFC algorithms.  The GSFC-S code is implemented using IDL functions and procedures, and the GSFC-M implementation uses both Mathematica and Python.

\subsubsection{Photon Trajectories and Time Delays}

In the GSFC codes the angular deflection in Equation~(\ref{eq:defl}) is computed using a change of variables to re-write the integrand (see \citealt{2002ApJ...564..353N}) using
$\hat b \equiv b/b_{\rm max} = \sin\alpha$ and $u\equiv R_S/R$. Then $\psi$ is evaluated numerically for a pre-defined grid of $\hat b$ values. As this essentially determines $\cos\alpha$ as a function of $\cos\psi$, the required derivative can also be evaluated using a three point Lagrangian interpolation scheme. The density of points in the $\hat b$ grid is increased as $\hat b$ approaches 1 to better resolve large bending angles (for example, $\hat b$ approaches 1 as an emitting element approaches eclipse), and also because as ${\hat b}\rightarrow 1$ the derivative term increases sharply.

Similarly, the time delay (Equation~\ref{eq:delay}) between a photon emitted with impact parameter $\hat b$ and one with $\hat b = 0$ is computed from a modified integral.
This gives the delay in units of $R/c$. This delay is computed
numerically using the same grid of $\hat b$ values as for the bending
angles.


In the GSFC codes the neutron star surface is discretized into area elements by defining $N_{\theta}$ and $N_{\phi}$ angular bins in the colatitude $\theta$ and azimuthal angle $\phi$, respectively. Given the hot spot's angular size and location (which at present are assumed to be circular), it is then determined which surface elements are part of the hot spot.  Only those elements are used to calculate the observed flux at any particular rotational phase.  

The first step is to choose the rotational phases of the star at which the observed flux is to be evaluated.  Then the flux contributions for all surface elements comprising the spot are computed.  The GSFC codes solve the equation
\begin{equation}
\phi_{\rm emit} + \Omega\Delta t(\hat b(\phi_{\rm emit})) = \phi_{\rm obs}
\end{equation} 
to find $\phi_{\rm emit}$ given $\phi_{\rm obs}$. In practice the GSFC codes use the computed time delays to construct this mapping as a look-up table for each surface element and each value of $\theta$ required for a calculation.  

This provides all the information necessary to compute the flux.  As part of the calculation it is necessary to specify a grid of observer energies at which the flux should be evaluated.  For each surface element contributing to the flux it is straightforward to use the relation between $E'$ and $E$ to convert the observed photon energies to ``local'' energies and then evaluate the intensity at those values.  The flux is then summed from all visible surface area elements at each rotational phase. In doing this the code takes advantage of the fact that the rotational waveform produced by an emitting element at a given co-latitude only needs to be computed once. Emission from a finite spot at a particular phase can then be computed by simply rotating the 
waveforms of the sub-elements comprising the spot to the correct phase.  The output of the calculation is the observed spectral flux at the selected rotational phases.


The current GSFC-S code is developed from a S+D code used to produce the results presented in \citet{2004AIPC..714..245S}, which was one of the first efforts to explore the quality of constraints on neutron star mass and radius that could be achieved by pulse waveform fitting with a much larger X-ray collecting area than the area of RXTE/PCA. That work was primarily in the context of X-ray burst oscillations, and as such the code allowed for additional effects related to X-ray burst phenomenology, such as spreading of the emitting hot spot. Additional recent results from the GSFC codes are presented in \citet{2016ApJ...818...93M}. 



The GSFC-M code has been developed in the last few years and some of the results from an earlier version of this code are presented in \citet{2016ApJ...818...93M} and \citet{2015arXiv150102776I}. This code has been written in Python. The deflection angle is computed using a grid in $\alpha$ from 0 to $\pi/2$ in steps of $\pi/2000$. All the interpolations in this code are quadratic. To compute the pulse waveforms with the desired accuracy of better than 0.1\%, the spot has been sampled at $N_{\theta}=100$ equally spaced latitudes, and $N_{\phi}=500$ equally spaced longitudes for each latitude, and the total flux is computed by summing all the flux contributions from each surface element in the emitting region.  

\subsection{The Illinois-Maryland code}

The Illinois-Maryland algorithm for computing waveforms using the Schwarzschild+Doppler (S+D) and oblate-star Schwarzschild-spacetime (OS) approximations is described in detail in a series of papers (see, e.g., Section~2 of \citealt{1998ApJ...499L..37M}, which introduced the S+D approximation; Sections~2.2 and 2.3 of \citealt{2009ApJ...706..417L}; Sections~2.1, 3.1, 3.3.1, and 4.1.2 of \citealt{2013ApJ...776...19L}; and Sections~2.1.3, 2.2, and 3.1 of \citealt{2015ApJ...808...31M}). In both approximations, the external spacetime is assumed to be the Schwarzschild spacetime. 


The Illinois-Maryland algorithm is fast and accurate: for example, a version that generates energy-resolved periodic pulse shapes for all the $E_{\rm obs}=1$~keV tests that agree to better than 0.1\% with the most accurate pulse shapes computed by the code validation team, at all pulse phases other than those extremely close to the flux minimum, takes $\approx 0.3$~seconds to compute the pulse shape on a single processor. As noted above, it is the speed with which a given algorithm can achieve the required accuracy that is important, rather than the order of its convergence. Convergence tests carried out using a previous version of the code showed rapid convergence to highly accurate pulse shapes (see Appendix~A of \citealt{2013ApJ...776...19L}, Sections A.1.1 and A.1.2). 

The pulse shape produced by radiation from a given emitting region is computed by sampling the region at $N_{\rm lat}$ equally spaced latitudes (typically $N_{\rm lat}=200$ is used). The axisymmetry of the Schwarzschild spacetime means that two photons that are emitted from the surface with the same values of $\theta,\ \alpha,\ {\rm and}\ \lambda$ 
but values of the stellar longitude that differ by $\Delta\phi$ will arrive at infinity at the same colatitude but separated in longitude by $\Delta\phi$. Thus, a ray needs to be traced only from a single stellar longitude at each relevant stellar latitude: rays from the other relevant longitudes can be generated by simply rotating the ray traced from the first longitude. A given emitting region is typically sampled at 100 equally spaced longitudes for each colatitude. The total observed flux from all the emitting regions can then be computed by summing the contributions, at the observer's colatitude and distance, from all the grid points within the emitting regions, for an adequate sample of pulse phases at the observer's colatitude and distance. The fluxes at any desired number $N_{\rm phase}$ of equally spaced pulse phases can then be determined by interpolating in the table of fluxes at the sampled pulse phases.  Quadratic or quartic interpolation is used in the Illinois-Maryland code depending on accuracy requirements.

The angular deflection of light rays from a static star and the propagation time relative to a radial ray given by the current code are in excellent agreement with the values computed using different algorithms by \citet{2013ApJ...776...19L} and the values computed using Mathematica routines (see \citealt{2013ApJ...776...19L}, sections A.1.1 and A.1.2). The angular deflection of a pencil beam from a small emitting spot on a rotating star is in excellent agreement with the value computed using a Mathematica routine (see \citealt{2013ApJ...776...19L}, section A.1.6).


\subsection{The Alberta Code}

The original versions of the Alberta S+D and OS codes were written in order to test these approximations against raytracing results in the background of a numerically generated \citep{1995ApJ...444..306S} relativistic rapidly rotating neutron star \citep{2007ApJ...654..458C}. In this early version of the code the OS approximation was implemented by embedding the exact oblate shape of a couple particular models in the Schwarzschild spacetime, and the main OS approximation given in Equation (\ref{eq:flux_os}) was introduced as an ansatz. 
\citet{2007ApJ...663.1244M} added a simple parametrization of the surface as well as a derivation of the OS approximation formulae presented in Section \ref{sec:OS}. These approximations were implemented in the code that was used to analyze data from three accreting ms-period X-ray pulsars \citep{2008ApJ...672.1119L,2009ApJ...691.1235L,2011ApJ...742...17L,2011ApJ...726...56M}. Many code improvements were implemented by \citet{2016ApJ...833..244S}, including the improved shape formula given in \citet{2014ApJ...791...78A}.

The Alberta code is written in C++ and makes use of libraries of routines from Numerical Recipes \citep{2002nrca.book.....P} and MATPACK\footnote{http://www.matpack.de}.
The implementation of this code is similar to the other codes described in this paper. The integrals for the deflection angles, lensing factors, and times of arrival are pre-computed for a fine two-dimensional grid of values of $M/R$ and $\alpha$. This grid is read into memory at the start of the waveform computation, similar to the Columbia and Illinois-Maryland codes.  

At each value of co-latitude covered by the spot, the value of $M/R$ is computed and the look-up tables for the deflection angle, lensing factor, and time of arrival are interpolated to create a set of one-dimensional tables for the correct value of $M/R$ for that latitude. Given a value of phase in the co-rotating frame of the star, the deflection angle required for the photon to be observed is computed.
If the deflection angle is less than or equal to the maximum allowed deflection angle (for the given $M/R$) the look-up tables are interpolated to find the required values of the zenith angle $\alpha$, the lensing factor, and the relative time of arrival. If the required value of the deflection angle is larger than allowed for a spherical star with the same value of $M/R$, then we check to see if it is possible for an initially ingoing photon to connect the star and the observer. If the required photon is initially ingoing, then the corrected values for the zenith angle and time of arrival are computed. Otherwise the photon is eclipsed and is not seen. 

These computations are repeated at regularly spaced values of the azimuthal angle on the star to create a waveform as seen in the observer's frame. Since the photons emitted at different values of phase take different amounts of time to travel to the observer, the resulting waveform is not evenly spaced in the observer's time. This waveform is interpolated to find the values of flux at regular time bins in the observer's frame. The interpolation is straight-forward, except in two cases. The waveform's maximum value (or similarly, a minimum) will typically fall between two time-bins in the irregularly spaced waveform. A parabolic interpolation scheme is used to correctly predict the location of the maximum and to predict the correct flux for the regularly spaced waveform. The other interesting case is when eclipses occur. In this case the non-zero values of flux near the eclipse are used to extrapolate the phases where the eclipse starts and ends. This then allows the flux near the eclipses to be correctly rebinned into regular time bins. The implementation of the interpolations near the maxima and eclipses has been tested by increasing the grid resolution in the azimuthal dimension to ensure that the rebinning step is accurate.

Once the observed waveform (at regularly spaced time bins) due to one infinitesimal spot on the star has been correctly created, the observed waveform for a spot at constant latitude and a range of azimuthal angles can be simply computed using the ``shift and add'' procedure used by the Columbia and Illinois-Maryland codes. The procedure is then repeated for the next value of latitude in order to compute a spot that spans a range of latitudes.

\subsection{The Amsterdam code}

The Amsterdam (AMS) light-curve code is integrated into the \textsl{X-ray Pulsation Simulation and Inference} (\textsl{X-PSI}) package\footnote{\url{https://github.com/ThomasEdwardRiley/xpsi}.} \citep{riley19b}, where it is called for numerical likelihood function evaluation. The first application of the Amsterdam light-curve code in a statistical analysis of X-ray data was by \cite{riley19} using \textsl{NICER} observations of PSR~J0030$+$0451. The basic algorithmic themes are described in sufficient detail elsewhere in this current paper. In this subsection we mention only novel aspects relevant to light-curve computation, and implementation-specific aspects that differ notably from the material outlined previously.

\textit{Surface discretization.}---A regular discrete representation of a radiating stellar surface is required for consistently fast likelihood function evaluation---an important consideration in large-scale statistical sampling applications. The surface is discretized with a regular mesh of points spaced in colatitude according to some criterion,\footnote{E.g., linear in colatitude, linear in cosine of colatitude, or by requiring that mesh elements enclose equal area on a Schwarzschild time-hyperslice.} and spaced linearly in azimuth about the stellar spin axis. These points have associated areas which weight in summation, the differential signals generated by material in their local vicinities. A mesh constructed from surface meridians and parallels in this manner leads to (infinitesimal) radiating elements mapping to one another via natural rotation of the star. This means that the exact signal (incident on a distant observer) generated by one element is related to the signals generated by a subset of other elements purely by time-translation (see, e.g., Appendix~\ref{app: CU code}), provided that the local comoving radiation fields within those elements are identical. If radiating elements of finite areal extent are identical under rotation, this time-translation symmetry holds exactly.  If such elements do not rotate onto each other exactly, the signals are, in approximation (see above), related by a factor equal to the ratio of areas. The representative points which generate the differential signals are the area-weighted mean points within the elements (before consideration of the boundary of the closed radiating region).

Admitting this symmetry, however, means that one is subject to a mesh which does not conform naturally to closed radiating regions whose boundaries are not constructed from coordinate isocurves. Moreover, the coordinate singularity at the poles means that the element shapes are far from congruent unless one allows element areas to span a wide range. On the other hand, having the element boundaries trace coordinate isocurves is natural from the perspective of integrating areas within closed regions on a rotationally deformed oblate surface, because the form of the differential area element is azimuthally invariant. Closed radiating regions will generally not conform exactly to some union of elements with trivially known area. It is therefore advisable to calculate the area of the radiating subset of each element.  This ensures that the total area of the radiating region is exact, and that the weights applied to the differential signals are more accurate. The AMS implementation efficiently assigns exact areas for several classes of radiating region (see \citealt{riley19}), but accuracy of course remains subject to issues such as spatial resolution and construction of non-congruent elements. The iso-latitudinal mesh offered by \citet{Gorski2005}, for example, constructs nearly congruent elements of equal area and would be close to ideal. However the element boundaries are more involved to handle when calculating the area of a radiating subset of an element. Nevertheless, this is a potential avenue for future improvement.


\textit{Ray-tracing.}---The rays are \textit{not} precomputed and written to disk for some discrete set of values of the dimensionless coordinate $r/M$. Instead, for each set of iso-colatitude elements (which by definition share a radial coordinate value) a set of rays is computed and stored in memory (for every likelihood function call). The cosine of the local ray angle, $\cos\alpha$, is linearly spaced between the local minimum (as permitted by surface tilt) and unity. We can afford this computation because it is far from being a bottleneck. One-dimensional spline interpolation of both local and integral quantities ($\cos\alpha$, ray lag, and also the ray bundle lensing factor via spline differentiation) is performed with respect to the cosine of the ray deflection, $\cos\psi$, as usual. The integral quantities (ray deflection, lag, and bundle lensing factor) are computed using a variable transformation. The point of this variable transformation is to eliminate the integrand divergence as $r\to r_{c}^{+}$,\footnote{We use an exact closed-form solution for $r_{c}$, that is equivalent to the expression given by \citet{Salmi18}.} and make the integral domain compact. Therefore a transformation $w=\sqrt{1-r_{c}/r}$ is implemented for geodesics whose $r_{c}$ exists, and $w=\sqrt{1-R/r}$ for those whose $r_{c}$ is undefined (plunging geodesics). The transformation is thus unique to every null geodesic\footnote{Whose spatial trajectory exists in a coordinate 2-plane through the stellar origin.} whose $r_{c}=r_{c}(R/M,\sin\alpha)$ exists and is labelled by a unique $\sin\alpha$.

\textit{Image-plane.}---For a number of the light-curve calibration exercises\footnote{These exercises were more involved than those explicitly presented in this paper, e.g., involving a numerical atmosphere model and two surface hot regions.} performed by this \textsl{NICER} working group, we internally cross-checked the AMS code with a code for integration via image-plane discretization.  The calculations were consistent to well within the accuracy threshold targeted in this paper.  The largest discrepancies occurrred as usual in the near vicinity of light-curve zero due to difficulty in accurately resolving images at the visible limb without extreme resolution.\footnote{The AMS image-plane code is relatively unsophisticated in comparison to modern open-source general purpose codes, some of which can operate in arbitrary spacetimes and coordinates, and on GPUs. GPUs are a more appropriate hardware choice due to the embarrassing parallelism of numerical integration of rays (generally coupled second-order non-linear ODE systems) backwards in time from the image-plane.} Image-plane discretization is inherently more expensive than surface discretization (and thus may not be tractable for use in large-scale sampling applications), but can be implemented in general purpose light-curve integrators. We applied the AMS variant by embedding the relevant oblate surface in a quasi-Kerr ambient spacetime \citep[][and references therein]{2014ApJ...792...87P} and then simply considering the limit of zero spin to recover the spacetime solution considered in this paper.


\section{Results of Comparison with Analytical and Semi-analytical Results}
\label{app:SDanalytical}

\subsection{Luminosity of a rotating, non-gravitating star}\label{sec:analytic}

The first semi-analytic code test is the computation of the bolometric luminosity of a rotating spherical Newtonian star including the effects of special relativity. For this computation, we assume a uniform temperature blackbody emitter as measured in the co-rotating frame, and a spherical star as viewed by a static observer. 

In order to compute the surface flux from a small surface element on the star's surface, it is useful to introduce a spherical coordinate system defined by the surface element's velocity vector, which is always perpendicular to the surface normal. The direction of any light ray, $\vec{k}$, emitted at the surface is then defined by the co-latitude $\xi$ measured from the velocity vector, and an azimuthal angle $\iota$ around the velocity vector. The definition of $\xi$ is such that it is the angle between the light ray and the velocity vector, agreeing with the definition in Section \ref{sec:SD}, and ranges from 0 to $\pi$. The azimuthal angle $\iota$ is defined so that $0\le\iota\le\pi$ corresponds to emission away from the surface, while $\pi\le\iota\le 2\pi$ corresponds to emission into the surface. With these definitions, the angle between the light ray and the surface normal, $\alpha$ is defined by $\cos\alpha = \sin\xi\sin\iota$. 

The frequency-integrated specific intensity of a blackbody is $I={1\over\pi}\sigma_{\rm SB}T^4$, where $T$ is the temperature and $\sigma_{\rm SB}$ is the Stefan-Boltzmann constant.  We assume that the blackbody temperature $T$ is the same everywhere on the surface of the star, as measured in the local comoving frame. 
Because $I/E^4$ is constant along rays, the frequency-integrated specific intensity of light emitted at a location ($\theta,\phi$) on the star is
\begin{equation}
I(\xi,\theta,\phi)={1\over\pi}\sigma_{\rm SB}T^4\left[{1\over{\gamma(\theta)(1-\beta(\theta)\cos\xi)}}\right]^4\; 
\end{equation}
in the static frame, where $\beta$ and $\gamma$ are defined by Equations (\ref{eq:beta}) and (\ref{eq:gamma}). 

The bolometric surface flux, $dF$ of the infinitesimal surface element is the normal projection of the specific intensity integrated over all angles $\xi,\iota$ that correspond to light emerging from the star,
\begin{equation}
dF(\theta,\phi)=\int_0^\pi \int_0^\pi I(\xi,\theta,\phi)\cos\alpha\sin\xi {\rm d}\xi {\rm d}\iota\; 
=
\int_0^\pi\int_0^\pi {1\over\pi}\sigma_{\rm SB}T^4\left[{1\over{\gamma(\theta)(1-\beta(\theta)\cos\xi)}}\right]^4\sin^2\xi \sin\iota {\rm d}\xi {\rm d}\iota\; .
\end{equation}

The total luminosity of the star is the integral of the surface flux over the surface,
\begin{equation}
L=\int_0^{2\pi}{\rm d}\phi 
\int_0^\pi \sin\iota {\rm d}\iota 
\int_0^\pi {\rm d}\theta \sin\theta 
\int_0^\pi {1\over\pi}\sigma_{\rm SB}T^4\left[{1\over{\gamma(\theta)(1-\beta(\theta)\cos\xi)}}\right]^4R^2\sin^2\xi {\rm d}\xi\; .
\end{equation}
Since the integrand is independent of $\phi$ and $\iota$, and $\beta(\theta) = \beta_{\rm eq} \sin\theta$ this reduces to 
\begin{equation}
L=4\pi\sigma_{\rm SB}R^2T^4\int_0^\pi \int_0^\pi {1\over\pi} \left[{1\over{\gamma(\theta)(1-\beta_{\rm eq}\sin\theta\cos\xi)}}\right]^4\sin^2\xi\sin\theta {\rm d}\xi {\rm d}\theta\; .
\end{equation}
When the rotation of the star vanishes, the integrand reduces to unity. For spinning stars, the lowest order special relativistic corrections increase the luminosity by an amount proportional to $\beta^2$.

As a specific example, suppose that a star rotating at an angular frequency of $\Omega=2\pi\times 600$~rad~s$^{-1}$ has a radius of $R=12$~km and the blackbody has a temperature in the comoving frame that is given by $kT=0.35$~keV.  Then this formula predicts that the luminosity will be $L=2.81381\times 10^{35}$~erg~s$^{-1}$.  One of our general pulse waveform codes finds $L=2.81323\times 10^{35}$~erg~s$^{-1}$, for a fractional accuracy of $2\times 10^{-4}$.

\subsection{Flux from a uniformly emitting, nonrotating star in general relativity}

The observed spectral flux from a uniformly emitting star in general relativity can be found by integrating the spectral flux given in equation (\ref{eq:flux}) over the visible parts of the star. If the star is non-rotating and the emission is isotropic the result is simply
\begin{equation}
F(E) =  (1-R_S/R)^{3/2} I(E') \int {\rm d}\Omega
\end{equation}
where $\int d\Omega$ is the solid angle subtended by the whole star. A spherical non-rotating star subtends an angle of
(e.g., \citealt{2013ApJ...776...19L} equation (A9), or \citealt{1993ApJ...413L..43M} equation (3))
\begin{equation}
\int {\rm d}\Omega=\pi\left(R\over D\right)^2(1-R_S/R)^{-1}\; .
\end{equation}

The parameters we consider for this test are the same as for the SD1 tests: $M=1.4$~\msol, $R=12$~km , 
$kT=0.35$~keV, and $D=200$~pc.  We again assume isotropic blackbody emission from each emitting point, as seen by a comoving observer, and using the codes discussed in Section~\ref{sec:verification} to numerically determine the expected counts per area per time per keV at an observed energy of 1~keV and compare them against the expected analytical result.
Plugging in the numbers for our example, we expect an unvarying 17.2279 counts~cm$^{-2}$~s$^{-1}$~keV$^{-1}$, and this is the average that we get to 1 part in $10^5$.

\subsection{Flux from a small spot on a non-rotating, gravitating star}
\label{app:nrstar}
We can use the result for the uniformly emitting star to predict the counts\,cm$^{-2}$\,s$^{-1}$\,keV$^{-1}$ expected from a uniform circular spot of angular radius $\Delta\theta$ at the moment that its center is directly underneath us, assuming that it is on a very slowly rotating star.  As before, the rate will be directly proportional to the solid angle that we see the spot subtend.  The angular radius of the spot as we see it is proportional to the specific angular momentum of the photons that we see from the edge of the spot, and in turn the specific angular momentum is proportional to the sine of the angle $\alpha$ made by the photon to the local surface normal as seen by a local comoving observer (because the star is spherical in these tests).  Thus if we find that an angle $\alpha$ from the surface normal produces a total deflection $\Delta\theta$ to infinity, then the flux we see from a uniform circular spot of angular radius $\Delta\theta$ will be $\sin^2\alpha$ times the flux from the full star.  For $\Delta\theta\ll 1$, $\alpha$ is very close to $\Delta\theta/(1+z)$, but for large $\Delta\theta$ numerical integration is required to determine $\alpha$.

For our parameters, $\Delta\theta=0.01$~rad means $\alpha=0.0080960007$~rad and we expect 0.00112918 counts~cm$^{-2}$~s$^{-1}$~keV$^{-1}$.  We see 0.00112919, for a fractional error of $8.9\times 10^{-6}$.  Similarly, $\Delta\theta=1$~rad means $\alpha=0.797364165$~rad and we expect 8.820079 counts~cm$^{-2}$~s$^{-1}$~keV$^{-1}$.  We see 8.8202, for a fractional error of $1.4\times 10^{-5}$.  

\subsection{Flux from a small spot on a rotating, gravitating star}

We will continue from the previous example by considering an observer on the rotational equator, and a spot with its center on the rotational equator.  Let the angular frequency seen by a distant observer be $\Omega$, which means that the angular frequency seen in a static frame just above the surface is $\Omega(1+z_{\rm grav})$.  Suppose that we are interested in the emission when the spot's center is at a phase $\Delta\theta$ before the spot is directly beneath us (i.e., the spot has a projected motion towards us).  The photons therefore need to have a total deflection angle $\Delta\theta$ to reach us.  This will be achieved with an angle to the surface normal, as measured in the local {\it static} frame, of $\alpha$.  This ray will have a propagation time to the distant observer that differs from the propagation time of a radial ray (again, in the static frame) by $\Delta t(\alpha)$.  Thus, as we discussed earlier, the phase of arrival at the observer, relative to the phase of arrival of a radial ray emitted when the spot is directly below the observer, is
\begin{equation}
\Delta\phi=-\Delta\theta+\Omega\Delta t(\alpha)\; .
\end{equation}

When we compare the number flux per unit frequency of such a spot with the number flux per unit frequency of a spot of the same size directly beneath us on a very slowly rotating star, we need to consider that the emitted photon energy will be changed by a factor $1+z_{\rm Dopp}$ compared with the emitted photon energy from a slowly rotating star, and thus the specific intensity will also be multiplied by $(1+z_{\rm Dopp})^{-3}$.

The observer and spot are both on the rotational equator, meaning that the angle $\xi$ between the photon direction and the direction of motion is $\xi=\pi/2-\alpha$, and similarly $\xi^\prime=\pi/2-\alpha^\prime$, where an unprimed angle is measured in the local static frame and a primed angle is measured in the local surface comoving frame.  As before, the Doppler shift is given by
\begin{equation}
1+z_{\rm Dopp}=\gamma(1-\beta\cos\xi)=\gamma(1-\beta\sin\alpha)\; .
\end{equation}
The zenith angle in the comoving frame, $\alpha'$, is given by the standard transformation (\ref{eq:alphatrans}), which reduces in this case to
\begin{equation}
\cos\alpha^\prime = \frac{\cos\alpha}{\gamma(1-\beta\sin\alpha)}.
\end{equation}


The flux measured at energy $E$ from the surface area element $dS^\prime$,
given by Equation (\ref{eq:flux}) is, for blackbody emission,
\begin{equation}
    dF(E)= \left(\frac{2}{c^2h^2} \frac{R^2}{D^2}(1-R_S/R)^{1/2} \frac{\diff\cos\alpha}{\diff\cos\psi} \sin\theta d\theta d\phi
    \right)
    \times \left( 
    \gamma \cos\alpha^\prime{1\over{e^{E'/kT}-1}}
    \right) \;
    \label{eq:example}
\end{equation}
where the first group of terms are independent of the star's spin, while the second group of terms depend on the spin. The factor of $\gamma$ comes from the definition of the surface area element in the rotating frame.


As a simple example, consider the moment when the spot is directly below the observer, corresponding to $\alpha=0$. For this case, $\cos\alpha^\prime=\gamma^{-1}$, cancelling out the $\gamma$ factor appearing in the second term of Equation (\ref{eq:example}). As a result, the ratio of flux for two stars with different rotation rates is just
\begin{equation}
\frac{dF(E_\Omega)}{dF(E_{\rm s})} = \frac{(e^{E'_{\rm s}/kT}-1)}{(e^{E'_\Omega/kT}-1)}.  
\end{equation}
where $dF(E_\Omega)$ is the flux from the surface element on the rotating star and $dF(E_{\rm s})$ is the flux from an identical static star. 
For the case of a star spinning at $200$ Hz (as measured by the observer)
this ratio for the measured flux at 1 keV is $(0.03000653/0.030218427)$, which can be multiplied by the flux from a small spot on a non-rotating star calculated in Appendix \ref{app:nrstar} to find the flux from a rotating star. For this example, we expect the exact result for the rotating star's flux at 1 keV to be $0.0011213$. We see $0.0011214$, for a fractional accuracy of about $10^{-4}$.



\acknowledgments
We thank J.~N\"attil\"a, F.~\"Ozel, D.~Psaltis, and J.~Poutanen  for insightful discussions and code comparisons. This work was supported in part by NASA through the \textsl{NICER} mission and the Astrophysics Explorers Program. M.C.M. is grateful for the hospitality of the Kavli Institute for Theoretical Physics at the University of California, Santa Barbara, during part of the writing of this paper, and was therefore supported in part by
the National Science Foundation under Grant No.\ NSF PHY-1748958.  M.C.M. is also grateful for the hospitality of the Perimeter Institute where part of this work was carried out. Research at Perimeter Institute is supported in part by the Government of Canada through the Department of Innovation, Science and Economic Development Canada and by the Province of Ontario through the Ministry of Economic Development, Job Creation and Trade. A.L.W. and T.E.R. acknowledge support from ERC Starting Grant No.~639217 CSINEUTRONSTAR (PI Watts). S.M.M. thanks NSERC for research funding. This research has made extensive use of NASA's Astrophysics Data System Bibliographic Services (ADS) and the arXiv.


\software{MATPACK~(\url{http://www.matpack.de}), Python/C~language~\citep{python2007}, GNU~Scientific~Library~\citep[GSL;][]{Gough:2009}, NumPy~\citep{Numpy2011}, Cython~\citep{cython2011}}

\bibliographystyle{aasjournal}
\bibliography{nicer_pulse_profile_modeling}

\listofchanges

\end{document}